\documentclass[
 reprint,
 superscriptaddress,
 preprintnumbers,
 amsmath,
 amssymb,
 prb,
 floatfix,
 aps,
]{revtex4-2}

\usepackage[utf8]{inputenc}
\usepackage[final]{changes}
\usepackage[english]{babel}
\usepackage[T1]{fontenc}
\usepackage{amsfonts}
\usepackage{amssymb}
\usepackage[normalem]{ulem}
\usepackage{graphicx}
\usepackage{braket}
\usepackage{verbatim}
\usepackage{hyperref}
\usepackage{subfigure}
\usepackage{longtable}
\usepackage{tabularx}
\usepackage{dcolumn}%
\usepackage{color}

\newcommand{\e}{\ensuremath{\mathrm{e}}}
\newcommand{\iu}{\ensuremath{\mathrm{i}}}

\newcommand{\dd}{\mathrm{d}}

\begin{document}

\author{Stefano Paolo Villani}
\email[]{email}
\affiliation{Dipartimento di Fisica, Università di Roma La Sapienza, Piazzale Aldo Moro 5, 00185 Roma, Italy}

\author{Lorenzo Monacelli}
\affiliation{Dipartimento di Fisica, Università di Roma La Sapienza, Piazzale Aldo Moro 5, 00185 Roma, Italy}

\author{Paolo Barone}
\affiliation{CNR-SPIN Institute for Superconducting and other Innovative Materials and Devices, Area della Ricerca di Tor Vergata, Via del Fosso del Cavaliere 100, I-00133 Rome, Italy}

\author{Francesco Mauri}
\affiliation{Dipartimento di Fisica, Università di Roma La Sapienza, Piazzale Aldo Moro 5, 00185 Roma, Italy}

\title{Role of ionic quantum-anharmonic fluctuations on the bond length alternation and giant piezoelectricity of conjugated polymers} 

\begin{abstract} 
Functionalized conjugated polymers are promising materials for electromechanical applications due to predicted giant piezoelectricity, arising from anomalously large dynamical effective charges and an enhanced response in the proximity of the dimerization phase transition. In this work, we assess the impact of quantum ionic fluctuations on piezoelectricity using the stochastic self-consistent harmonic approximation with a Rice-Mele diatomic chain model, parametrized to reproduce hybrid-functional first-principles calculations of prototypical carbyne. The model's accuracy is validated against first-principles calculations both with and without quantum-anharmonic effects. We find that ionic fluctuations strongly impact the structural properties, with the boundary of the dimerization phase transition shifted by $34\%$. Despite quantum fluctuations in the bond length reaching magnitudes comparable to the average, the strong piezoelectric response persists. The topological enhancement of the effective charges remains robust and is even enhanced by about $~20\%$ thanks to a quantum-induced shrinking of the electronic gap. The piezoelectric coefficient remains dominated by the internal relaxation and retains a morphotropic-like character, reaching maximum values near the renormalized boundary, with quantum anharmonicity mainly shifting the optimal enhancement window.
\end{abstract}

\maketitle 
\section{Introduction} 
Conjugated polymers (CPs) are organic materials characterized by a backbone chain of carbon atoms, whose overlapping $\pi$-orbitals result in an effective 1D system of delocalized electrons along the chain. 
Linear CPs often display an ordered dimerized structure with bond-length alternation (BLA) of short and long bonds, whose intimate relationship with non-trivial electronic properties and topological features has been studied for many decades\cite{su1979solitons,heeger2001,barford2013electronic}.
Thanks to their multifaceted theoretical and technological appeal, the interest in CPs has grown since their discovery to embrace many different fields. Indeed, the high responsiveness of the delocalized electrons underlies a variety of 
peculiar electro-optical properties that make CPs functional materials for a wide range of technological applications
as organic semiconductors\cite{fratini2020charge,guo2013designing,moliton2004review}, organic solar panels\cite{chueh2019recent,gunes2007conjugated}, organic light-emitting diodes\cite{song2020organic,alsalhi2011recent}, organic field-effects transistors\cite{bao2018organic,muccini2006bright,horowitz1998organic} and for bioelectronics\cite{zeglio2019conjugated,inal2018conjugated}. 

More recently, a strategy has been proposed to enhance the piezoelectric response of functionalized CPs\cite{villani2024giant}, possibly opening a promising avenue for electro-mechanical applications\cite{tadigadapa2009piezoelectric,briscoe2015piezoelectric} using organic piezoelectrics that combine high mechanical flexibility, low fabrication costs and biocompatibility\cite{setter2006ferroelectric,lovinger1983ferroelectric,ramadan2014review,bowen2022advfuncmat,PVDF_review2020} hardly attained in inorganic piezoelectric ceramics\cite{berlincourt1971piezoelectric,jaffe1958piezoelectric,damjanovic2009review}. The predicted enhancement is rooted in the adiabatic Thouless pump\cite{thouless1983quantization}, which may give rise to an electronic polarization of topological and nonlocal origin when inversion symmetry is broken\cite{onodaPRL2004, yamauchi2014electronic}. A small atomic displacement is then responsible for a huge polarization change, a dynamical quantity known as Born effective charge, that measures the lattice dielectric response to infrared light\cite{born1968} as well as the internal-relaxation contribution to the piezoelectric response\cite{piezoMartin1972}. An arbitrarily strong piezoelectricity can be further attained if the functionalized CP is pushed close to the boundary of a second-order structural transition, taking advantage of the flattened free-energy profile and consequent diverging behaviour of the internal strain, akin to the enhancement of the piezoelectric response close to the morphotropic phase boundaries of ferroelectric oxides\cite{ahart2008origin,damjanovic2010morphotropic}. This can be achieved by changing the chemical composition of a CP, either by substitutional doping within the backbone chain or by introducing different functional groups bonded to carbon atoms. Both functionalization strategies introduce atomic inequivalence in the chain, allowing, on the one hand, for non-zero Born effective charges and, on the other hand, for piezoelectricity when combined with a BLA distortion, breaking inversion symmetry. At the same time, the presence of inequivalent monomers within the CP chain would open a gap in the electronic band structure even in the undimerized structure\cite{rice1982elementary}, weakening the tendency to display BLA and thus serving as a functional knob to control a morphotropic-like composition-driven phase transition. According to the predictions of Ref. \cite{villani2024giant}, the giant piezoelectric response in CPs is ultimately dominated by the internal-relaxation contribution, benefiting from the enhancement of both Born effective charges and internal strain close to the transition point and requiring the simultaneous presence of atomic inequivalence and BLA.

Bond-length alternation in CPs has been a subject of long-lasting debate, being strongly dependent on both electron-electron (e-e) and electron-phonon (e-ph) interactions. The conventional wisdom attributes its origin mostly to the tendency of metallic undimerized monoatomic chains to display a charge-density wave, triggered by a Fermi-surface nesting that enhances the density-density response at the nesting wave-vector\cite{peierls1996quantum}. For a free electron gas, an arbitrarily small coupling with the lattice may release the electronic instability by opening a gap through a periodic lattice distortion, thus inducing a metal-insulator transition accompanied by the BLA\cite{frohlich1954theory,pouget2016peierls}. Long-range electron correlation and the interplay between e-e and e-ph interactions may significantly affect the strength or even the onset of the BLA distortion\cite{elcorr_PRL1987,comment1989_elcorr_PRL1987}. Indeed, predicting accurate BLA has long been considered a necessary benchmark for assessing the accuracy of different electronic structure methods\cite{jacquemin2005assessment, Choi_jcp1997_BLA_elstru, Chabbal_hybrid2010, Adamo_jctc2011_dftbenchmarks, ferretti2012ab}, with the general consensus that, within the framework of Density Functional Theory (DFT), corrections beyond local and semilocal approaches are required. On the other hand, the long-range dimerization order in simple models of monoatomic chains has been shown to be robust with respect to quantum lattice fluctuations at $T=0$, albeit with a significant renormalization of the BLA \cite{Hirsch_prb1983_qf_ssh, Hanke_prb1986_qf_ssh, Barford_prb2006_qf_ssh, Bakrim_prb2007_qf_ssh}. A recent first-principles computational study of carbyne -- a prototypical CP chain comprising carbon atoms only-- including both e-e interaction and quantum-anharmonic effects (QAE) with post-DFT approaches confirmed the robustness of the dimerized polyynic phase over the undimerized cumulenic one, further unveiling an increasing role of QAE with increasing temperature and questioning the Landau-Peierls picture for the temperature-driven second-order structural phase transition\cite{romanin2021dominant}. 

As anticipated, the introduction of a site-dependent modulated potential modelling the atomic inequivalence along the chain would remove the electronic instability by opening a gap in the band structure, thus ruling out the Peierls instability. The stabilization of a dimerized phase stems instead from the competition between the e-ph and on-site potential energy scales\cite{rice1982elementary}, where the latter would favour a site-centered charge-ordered insulating phase with no BLA. Neglecting lattice fluctuations and at $T=0$, a second-order phase transition to a dimerized structure with the BLA as the order parameter can be induced by decreasing the on-site potential (increasing the e-ph) at fixed e-ph interaction (on-site potential)\cite{villani2024giant}. 
However, the robustness of such phase transition with respect to QAE and quantum/thermal lattice fluctuations has yet to be assessed, given that strong anharmonicity -- naturally appearing close to any displacive phase transition -- is expected to be further enhanced in 1D systems\cite{cignarella2025extreme}. Similarly, the argued topological protection against quantum fluctuations and anharmonic effects of the giant Born effective charges predicted in Ref. \cite{villani2024giant} needs to be validated. 

Prompted by these reasons and aiming at verifying the robustness of the predicted giant piezoelectricity in CPs against quantum and thermal fluctuations, in this work we adopt a non-perturbative treatment of QAE, based on the self-consistent harmonic approximation\cite{Hooton_scha1955, Bowman_scha.1978} in its stochastic implementation (SSCHA)\cite{errea2014anharmonic,errea2016quantum,bianco2017second,monacelli2021black,monacelli2021stochastic,monacelli2025analyzing}. To address systematically the impact of QAE on the morphotropic-like transition in $\pi$-conjugated chains and to lessen the computational burden of DFT+SSCHA computations\cite{romanin2021dominant}, we consider a working model based on the well-known Rice-Mele diatomic chain model\cite{rice1982elementary}, with parameters chosen to reproduce hybrid-functional DFT calculations of carbyne and decorated carbyne at $T=0$. The predictive power of the model is benchmarked with respect to fully first-principles calculations of polar responses (effective charges and piezoelectric coefficients) as well as to quantum anharmonicity and thermal fluctuations by comparing with available results on carbyne\cite{romanin2021dominant}, quite unexpectedly yielding semi-quantitative agreement in both cases. Having assessed the reliability of the simplified model, we turn to discuss QAE on the stabilization of the dimerized phase and on the effective charges, providing insight into how quantum anharmonicity affects the morphotropic-like and topological contributions to the enhanced piezoelectric response. 

The paper is organized as follows. In Section \ref{sec1}, we introduce the general framework for studying QAE on functionalized CPs. After a brief review of SSCHA, this section is devoted to evaluating realistic parameters for the considered effective model and validating its predictive power. The model is then used to assess the effects of quantum anharmonicity and finite temperatures on the stabilization of the dimerized phase in Section \ref{sec2}, and on Born effective charges in Section \ref{sec3}. Section \ref{sec4} is devoted to the piezoelectric response, building on the insight provided by previous sections, while we draw our conclusions in Section \ref{sec5}.

\section{Theoretical framework} \label{sec1}
\subsection{The stochastic self-consistent harmonic approximation}
The stochastic self-consistent harmonic approximation has proven to be one of the most reliable approaches to treat quantum-anharmonic and thermal ionic fluctuations in solid-state systems \cite{errea2014anharmonic,errea2016quantum,bianco2017second,monacelli2021black,monacelli2021stochastic,monacelli2025analyzing}, even when other methods, e.g., path-integral molecular dynamics\cite{ceperley1995path}, fail or become too expensive.
The equilibrium quantum distribution is approximated by a trial Gaussian ionic density matrix, whose parameters are optimized to minimize the free energy. The variational, non-perturbative approach accounts for the quantum nature of the ions, simultaneously including the anharmonic terms of the Born–Oppenheimer potential to all perturbative orders as well as the effects of finite temperatures. The minimization procedure requires calculating the electronic energies and forces for many supercell configurations, a task performed by an external calculator coupled to the SSCHA code. Within the SSCHA framework, QAE are incorporated on a quantity $O(\boldsymbol{R})$ which depends on the atomic positions $\boldsymbol{R}$ by averaging the value $\langle O (\boldsymbol{R})\rangle_\rho$ over $N_\mathrm{conf}$ supercells configurations randomly generated from the Gaussian quantum density matrix $\hat\rho$ using a Monte Carlo approach: 
\begin{align} 
    \label{eq:QAE_average}
    \langle O(\boldsymbol{R}) \rangle_\rho 
    &= \int d\boldsymbol{R} O(\boldsymbol R) \rho(\boldsymbol{R}) \approx \frac{1}{N_\mathrm{conf}}\sum_{\mathcal{I}=1}^{N_\mathrm{conf}} O(\boldsymbol{R}_\mathcal{I}),
\end{align} 
where each supercell has a different atomic structural configuration $\boldsymbol{R}_\mathcal{I}$. Substantial evidence points to the use of range-separated hybrid functionals to account for the effects of the long-range electron-electron correlation of the delocalized electrons on the BLA of CPs\cite{jacquemin2005assessment, peach2007, ferretti2012ab,romanin2021dominant,romanin2022giant,barborini2022excitonic,villani2024giant}. However, free energy minimizations and average-value calculations with these functionals are extremely expensive, as noted, e.g., in Ref.\cite{romanin2021dominant}. For this reason, we constructed a working chain model, introduced and validated in the remainder of this section
to enable fast \textit{and} reliable calculations of electronic forces and energies as well as for an efficient inclusion of QAE on CPs properties. 

\subsection{Effective model for QAE in conjugated chains
} 
We describe the backbone chain of CPs as a 1D collection of $N_\mathrm{at}$ atoms. We indicate with $r_i$ the position of atom $i$ along the chain, with $\boldsymbol{R}=(r_1,\dots,r_{N_\mathrm{at}})$ the collection of all the atomic positions, and we consider periodic boundary conditions.  
We introduce an atomic electronic inequivalence along the chain via a site-dependent modulated potential $\Delta(r_i)$. Without loss of generality, we assume $\Delta(r_i)=(-1)^i\Delta$, where $\Delta>0$ is an onsite energy term, and we consider a reference chain of equidistant atoms with bond length $a/2$. This allows us to identify a unit cell  of length $a$ comprising two
neighbouring atoms with onsite energy $-\Delta$ and $+\Delta$, respectively, and define a supercell as the collection of $N_\mathrm{cells}$ adjacent elementary cells, with $N_\mathrm{at}=2 \times N_\mathrm{cells}$. If $\Delta=0$, neighbouring atoms are electronically equivalent and we recover the well-known SSH model\cite{su1979solitons}, usually adopted to describe CPs such as carbyne or polyacetylene (PA), a CP made by the repetition of a C$_2$H$_2$ unit. If $\Delta\neq0$, instead, we recover the Rice-Mele diatomic chain model\cite{rice1982elementary}, originally introduced to describe
substituted polyacetylenes (SPA), a class of CPs formed by inequivalent monomers through substitutions within 
the C$_2$H$_2$ unit of PA
\cite{masuda2007substituted}.
We treat the electrons delocalized along the chain in a nearest-neighbour tight-binding approximation and indicate with $t_{i+1,i}$ the hopping integral between the $\pi-$orbitals of atoms $i$ and $i+1$. Aiming to study the manifestation of BLA,  
at linear order in the atomic displacements we write $t_{i+1,i} = t_0 -\beta \delta r_{i+1,i}$, 
where $t_0$ is the site-independent hopping energy of the chain with equidistant atoms, $\beta$ is an e-ph coupling parameter, and $\delta r_{i+1,i} = r_{i+1} - r_i - a/2$ quantifies neighbouring atoms' relative displacement with respect to their positions in the equally-spaced chain. For simplicity, we consider only longitudinal displacements, parallel to the linear-chain direction. 
The ion-ion interactions are effectively accounted for by an elastic-energy term with spring constant K, which favors the configuration with equidistant atoms. 
The 
Hamiltonian of the supercell reads:
\begin{align}
\label{eq:H_tot}
H_\mathrm{tot} = T_\mathrm{ion} + \sum_{i=1}^{N_\mathrm{at}} \Bigg[ 
    \frac{1}{2}\mathrm{K} \delta r_{i+1,i} ^2 + 2n_\mathrm{e}\Delta(-1)^ic^\dagger_ic^{\phantom\dagger}_i + \nonumber \\ - 2n_\mathrm{e} \left[\left( t_0 -\beta\delta r_{i+1,i}  \right) c^\dagger_{i+1} c^{\phantom\dagger}_i + \mathrm{h.c.} \right]
\Bigg],
\end{align}
where $T_\mathrm{ion}$ is the kinetic energy operator for the ions, the factor of $2$ accounts for the spin degeneracy, $n_\mathrm{e}$ is the number of electronic $\pi$-orbitals per carbon atom, with $n_\mathrm{e}=1$ for (S)PA and $n_\mathrm{e}=2$ for (decorated) carbyne, and $c^\dagger_i/c^{\phantom\dagger}_i$ are creation/annihilation operators for electrons. The free energy of the system is then defined following the SSCHA approach as  
\begin{equation}
    F[{\rho}](t_0,\beta,\mathrm{K},\Delta) = \langle E_\mathrm{tot}(\boldsymbol{R};t_0,\beta,\mathrm{K},\Delta) \rangle_{{\rho}} - k_\mathrm{B}TS[{\rho}]
\end{equation}
where ${\rho}(\boldsymbol{R})$ is the trial Gaussian density matrix to optimize to minimize the free energy, $\langle E_\mathrm{tot}(\boldsymbol{R};t_0,\beta,\mathrm{K},\Delta) \rangle_{{\rho}}$ is the average value of the BO energy $E_\mathrm{tot}$ of the system, obtained from $H_\mathrm{tot}$ as described in Appendix \ref{app:model_forces_energies}, $T$ is the temperature, $k_\mathrm{B}$ the Boltzmann constant and $S$ the entropy. 
The minimization procedure to obtain the optimal $\rho_\mathrm{min}$ requires electronic energies and forces, which we compute from $H_\mathrm{tot}$ (see Appendix \ref{app:model_forces_energies}). 

The ordered dimerized structure can be seen as a repetition of $N_\mathrm{cells}=N_\mathrm{at}/2$ diatomic cells displaying an alternation between a bond length $l_1$ between the two atoms in the same cell, and a bond length $l_2$, between an atom and its nearest neighbour in the adjacent cell. In this case, the BLA is simply quantified by the difference between $l_1$ and $l_2$, namely 
\begin{align}
    \label{eq:def_BLA_tm}
    \mathrm{BLA} \equiv |l_1 - l_2|. 
\end{align}
The value of this difference is the same for any unit cell $n$, and can be computed as
\begin{align}
    l_1 - l_2 
    &= (r_{2n} - r_{2n-1}) - (r_{2n+1} - r_{2n}) \nonumber \\
    &= 2(r_{2n} - r_{2n-1}) - a, \;\;\; \forall~n=1,\dots,N_\mathrm{cells}
\end{align}
where $r_{2n}$ and $r_{2n-1}$ are the positions of the atoms belonging to two different sublattices in the $n$-th unit cell, and we exploited the fact that $r_{2n+1}=r_{2n-1}+a$.
In the presence of fluctuations, all the pairs of neighbouring atoms now have different bond lengths. However, we assume that we can always distinguish between the bonds of atoms in the same cell and those of atoms in adjacent cells. In this way, following the prescription of Equation (\ref{eq:QAE_average}), we compute the value of the bond length difference corrected by quantum-anharmonic effects
by averaging their values over the supercell ionic configurations sampled by the quantum dynamics: 
\begin{equation}
    \label{eq:BLA_QAE}
    \langle l_1 - l_2\rangle_\rho \approx \frac{1}{N_\mathrm{conf}} \sum_{\mathcal{I}=1}^{N_\mathrm{conf}}\sum_{n=1}^{N_\mathrm{cells}} \frac{2(r_{2n,\mathcal{I}}-r_{2n-1,\mathcal{I}})}{N_\mathrm{cells}} - a,
\end{equation}

\subsection{Tuning model's parameters in the absence of fluctuations}
If we neglect the ionic fluctuations, the chain model reduces to the well-known Rice-Mele diatomic chain model\cite{rice1982elementary}.
The parameter $\Delta$ guides a second-order structural phase transition between a lower-symmetric phase with BLA, exhibiting non-zero effective charges and piezoelectricity, and a higher-symmetric phase, with no BLA and hence no electro-mechanical activity\cite{villani2024giant}. Despite its simplicity, the model captures the main features of CPs with few parameters that have a direct physical interpretation.
This prompted us to calibrate the values of the parameters of the model to reproduce target hybrid-functional DFT calculations on CPs in the absence of ionic quantum effects. Motivated by the results of Ref.\cite{romanin2021dominant}, we choose carbyne as an exemplary reference and, aiming to exploit the model as a reliable calculator for energies and forces, we considered target quantities derived from its total energy profile computed with hybrid functional PBE0\cite{adamo1999toward}.
From a diatomic unit-cell relaxation, we obtain the lattice constant $a=2.354$ \AA. Since carbyne comprises only equivalent C atoms, one has $\Delta=0$: in the next subsection we will validate the reliability of such modelization of atomic electronic inequivalence in a decorated carbyne system. The values of $t_0$, $\beta$, and $\mathrm{K}$ were then chosen to reproduce: (i) the value BLA$^\mathrm{DFT}_\mathrm{min}$ which minimizes the energy and characterizes the polyynic phase of carbyne; (ii) the energy gain $E^\mathrm{DFT}_\mathrm{gain}$ between the cumulenic phase, with no BLA, and the polyynic phase with BLA$^\mathrm{DFT}_\mathrm{min}$; (iii) the frequency at $\Gamma$ of the longitudinal optical phonon $\omega^\mathrm{DFT}_\mathrm{LO}(\Gamma)$ that drives the structural phase transition, which is related to the curvature of the energy profile in its minimum. The procedure for obtaining the values of the model's parameters is described in Appendix \ref{app:model_fit}, whereas in Table \ref{tab:fitted_parameters} we present our results, along with the values of the target quantities obtained from the DFT calculations with hyrbid functional PBE0. Computational details on the \textit{ab initio} calculations are provided in Appendix \ref{sec:app_comp_detail}. 
\begin{table}[htb!]
    \centering
    \begin{tabular}{ccc}
        \hline \hline 
        \rule{0pt}{1.2em} $t_{0}\;(\mathrm{eV})$ & $\beta\;(\mathrm{eV/}$\AA$)$ & $\mathrm{K}$ (eV/\AA$^2$) \\ 
        2.38455 & 7.20360 & 127.97659 \\ 
        \hline 
        \rule{0pt}{1.2em} $E^\mathrm{DFT}_{\mathrm{gain}}\;\mathrm{(meV/atom)}$ & BLA$^\mathrm{DFT}_\mathrm{min}$ (\AA) & $\omega^\mathrm{DFT}_{\mathrm{LO}}(\Gamma)\;(\mathrm{cm^{-1}})$ \\ 
        \rule[-.4em]{0pt}{0pt} 
        33.86570 & 0.10011 & 1986.57 \\ 
        \hline 
        \rule{0pt}{1.2em} $E_{\mathrm{gain}}\;\mathrm{(meV/atom)}$ & BLA$_\mathrm{min}$ (\AA) & $\omega_{\mathrm{LO}}(\Gamma)\;(\mathrm{cm^{-1}})$ \\
        33.86572 & 0.10010 & 1986.50 \\
        \hline \hline
        \vspace{0.2cm}
    \end{tabular}
        \caption{In the first line, values of the model's parameters tuned to reproduce \textit{ab initio} results for carbyne. In the second line we list the \textit{ab initio} DFT values of the targeted quantities computed with the PBE0 hybrid functional: $E^\mathrm{DFT}_{\mathrm{gain}}$ is the depth of the total energy profile of carbyne; BLA$^\mathrm{DFT}_\mathrm{min}$ is the bond length alternation which minimizes the total energy; $\omega^\mathrm{DFT}_{\mathrm{LO}}(\Gamma)$ is the frequency of the longitudinal optical phonon computed for the configuration with BLA$^\mathrm{DFT}_\mathrm{min}$. In the last line, values of the same quantities obtained using the model with the values of the parameters shown in the first line of the table. 
        }
    \label{tab:fitted_parameters}
\end{table}%

\subsection{Validation of the model in the absence of fluctuations} 
\begin{figure}[!htb]
    \begin{minipage}[c]{1.\linewidth}
    \centering 
    \includegraphics[width=\textwidth]{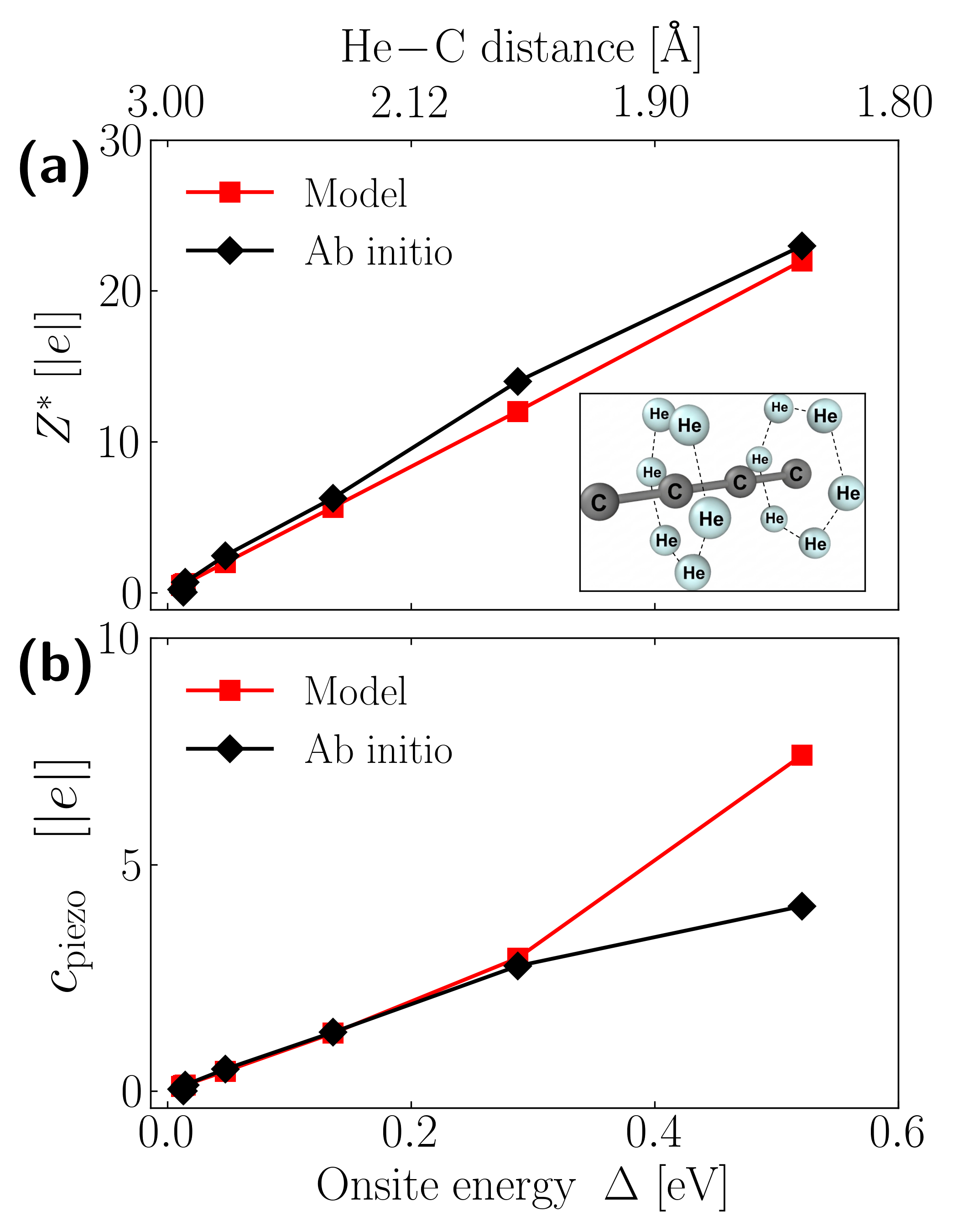} 
    \caption{
    Comparison of \textit{ab initio} values and model prediction of Born effective charge $Z^*$ \textbf{(a)} and piezoelectric coefficient $c_\mathrm{piezo}$ \textbf{(b)} evaluated at different He-C distances, shown on the horizontal axis at the top. To each value of the distance, it corresponds a fitted value of $\Delta$ in the model, shown on the horizontal axis at the bottom. In the inset, we illustrate the decorated carbyne system displaying two simulation cells. 
    }
    \label{fig:fit_Zeff_piezo}
    \end{minipage}
\end{figure}

As discussed in the introduction, a necessary condition for a CP to manifest a piezoelectric response is to have electronically inequivalent monomers along the chain. 
In the model, this requirement can be satisfied by considering a $\Delta\neq0$. As a realistic counterpart to this inversion-symmetry-breaking mechanism, we considered a decorated carbyne system, obtained by placing six helium atoms around one carbon atom every two along the backbone chain. Helium atoms were placed at the vertices of an equilateral hexagon perpendicular to the linear chain direction, with the carbon atom at its center. In this way, the distance between the vertices and the center, i.e., the He-C distance $d_{\mathrm{He- C}}$, acts as a knob to locally modify the electronic charge density of the delocalized electrons of carbyne by exploiting the interaction with the closed-shell orbitals of the helium atoms. Assuming that the values of model's parameters $t_0$, $\beta$ and $\mathrm{K}$ that we tuned for carbyne remain unchanged, each value of $d_{\mathrm{He-C}}$ corresponds to a value of $\Delta$ in the model, that we extracted using again the energy gain between the dimerized and undimerized structure of decorated carbyne and the optimized BLA in the dimerized structure. In the inset of Figure \ref{fig:fit_Zeff_piezo}(a), we show two simulations cells of the decorated carbyne system.
We report in Table \ref{tab:fit_Delta} the obtained values of $\Delta$, along with the corresponding values of $d_\mathrm{He-C}$, as well as the values of the target quantities obtained \textit{ab initio} and using the model. The details of the fitting procedure can be found in Appendix \ref{app:model_fit}. 

In order to validate the reliability and predicitive power of the simple modelization of the atomic electronic inequivalence via a single parameter $\Delta$, we compared the polar responses of the decorated carbyne computed at the PBE0 level with those obtained in the model. Specifically, we computed the values of the Born effective charges and of the piezoelectric coefficient in the model for each value of the fitted $\Delta$. These two quantities measure different responses of the electronic charge density and are defined as derivatives of the electronic polarization $P$, which in the absence of fluctuations is defined as the dipole moment of the diatomic cell divided by the cell length. The effective charge $Z^*_i$ of atom $i$ is then defined as the derivative of $P$ with respect to the displacement of atom $i$, namely
\begin{equation}
    \label{eq:zeff}
    Z^*_i = a\frac{\partial P}{\partial r_i}.
\end{equation}
The charge-neutrality acoustic sum rule implies $Z^*_i = (-1)^iZ^*$, allowing us to compute a single $Z^*$ using linear perturbation theory, as shown in Appendix \ref{app_Zeff}. 
The piezoelectric coefficient $c_\mathrm{piezo}$ quantifies the electro-mechanical response of the electronic charge density to a homogeneous strain $\epsilon$ and is defined as the derivative of $P$ with respect to $\epsilon$, namely
\begin{align}
    \label{eq:def_piezo}
    c_\mathrm{piezo} = \frac{\dd P}{\dd \epsilon}
\end{align}
that we compute in the model as finite differences of $P$, as detailed in Appendix \ref{app_piezo}. 
In Figure \ref{fig:fit_Zeff_piezo}, we compare the values of $Z^*$ and $c_\mathrm{piezo}$ obtained in the model for the different values of $\Delta$, with the values computed \textit{ab initio} in the corresponding carbyne-decorated system. The remarkable agreement we find validates the model's predictive power with respect to the Born effective charges and the piezoelectric coefficient. 
\begin{table}[htb]
    \centering
    \begin{tabular}{cccccc} 
    \hline \hline
    \rule{0pt}{1.2em} 
    $d_{\mathrm{He-C}}$ & $\Delta$ & BLA$_{\mathrm{min}}^{\mathrm{DFT}}$ & BLA$_{\mathrm{min}}$ & $E_{\mathrm{gain}}^\mathrm{DFT}$ & $E_{\mathrm{gain}}$ \\ 
    (\AA) & (eV) & (\AA) & (\AA) & (meV) & (meV) \rule[-3.0pt]{0pt}{0pt}  \\
    \hline\hline
    1.80 & 0.521 & 0.050 & 0.068 & 10.708 & 8.632 \\
    2.00 & 0.287 & 0.084 & 0.092 & 32.875 & 36.334 \\
    2.20 & 0.136 & 0.094 & 0.098 & 56.633 & 56.718 \\
    2.40 & 0.047 & 0.096 & 0.100 & 65.518 & 65.421 \\
    2.60 & 0.015 & 0.096 & 0.100 & 67.651 & 67.178 \\
    2.80 & 0.013 & 0.096 & 0.100 & 67.881 & 67.234 \\
    3.00 & 0.012 & 0.097 & 0.100 & 67.825 & 67.284 \\ 
    \hline \hline
    \end{tabular}
    \caption{
    Onsite energy $\Delta$ fitted for each value of the distance $d_{\mathrm{He-C}}$ between the carbon atom and the helium atoms in decorated carbyne, described in the text and shown in the inset of Figure \ref{fig:fit_Zeff_piezo}. Model parameters $t_0$, $\beta$ and $\mathrm{K}$ are kept fixed at values obtained for carbyne and given in Table \ref{tab:fitted_parameters}.    
    Values in the columns BLA$_{\mathrm{min}}^{\mathrm{DFT}}$ and $E_{\mathrm{gain}}^\mathrm{DFT}$ are reference values for the optimized BLA and for the energy gain of the dimerized structure of decorated carbyne obtained with PBE0 functional, to be compared
    with corresponding values 
    computed in the model with the fitted value of $\Delta$.}
    \label{tab:fit_Delta} 
\end{table} 

\subsection{Validation of the model in the presence of quantum-anharmonic and thermal fluctuations} 
\begin{figure}[!htb]
    \begin{minipage}[c]{1.0\linewidth}
    \centering
    \includegraphics[width=\linewidth]{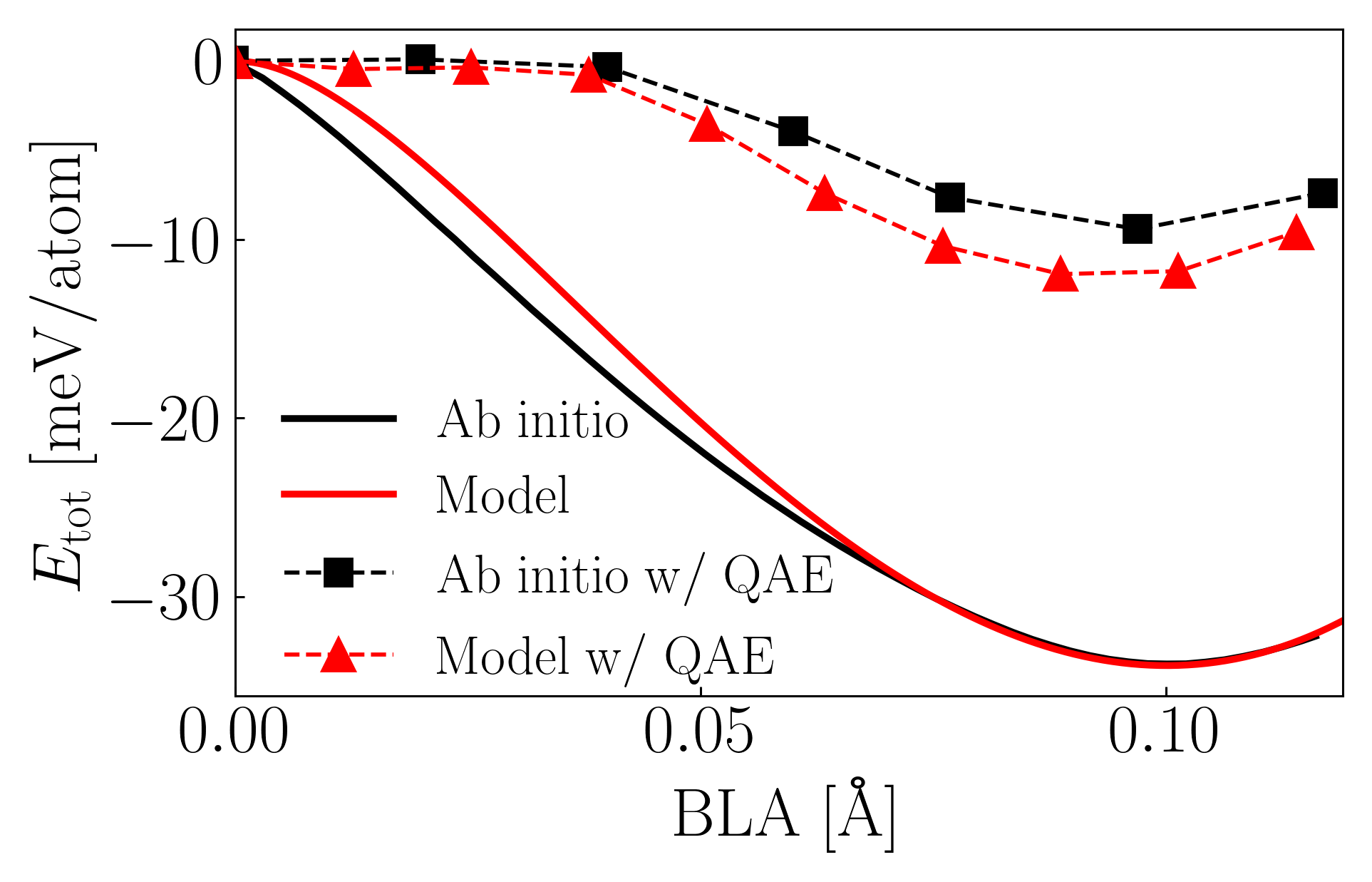}
    \caption{ 
    Energy profile of carbyne as a function of the bond-length difference BLA with and without quantum-anharmonic effects (dots and lines, respectively). Fully ab initio results obtained with PBE0 functional and SSCHA are shown in black. Model results, obtained using parameters of table \ref{tab:fitted_parameters}, are shown in red.
    }
    \label{fig:comparison_Etot_QAE}
    \end{minipage}
\end{figure}

Finally, we validate the accuracy of the model in describing the effects of ionic fluctuations in carbyne. First, we benchmark the $T=0~\mathrm{K}$ quantum-anharmonic energy profile obtained using the model with $\Delta=0$, against the one obtained in Ref.\cite{romanin2021dominant} with hybrid-functional DFT + SSCHA calculations. As shown in Figure \ref{fig:comparison_Etot_QAE}, quantum-anharmonicity in the model results in a reduction of the energy difference between the polyyne and cumulene phases of carbyne of $\sim65\%$, in remarkable agreement with the reference value of $\sim70\%$. 

\begin{figure}[!htb]
    \begin{minipage}[c]{1.0\linewidth}
    \centering
    \includegraphics[width=\textwidth]{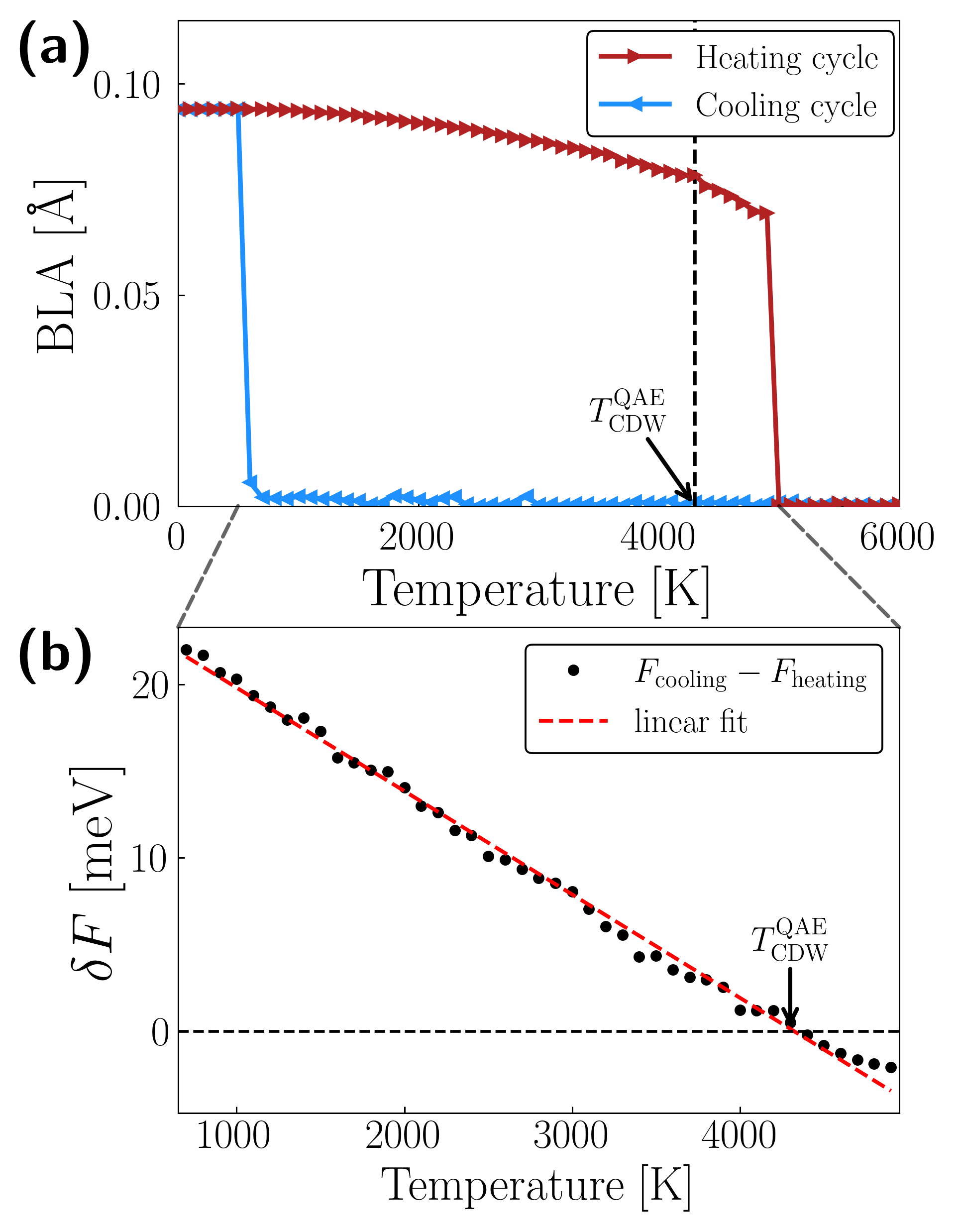}
    \end{minipage}
    \caption{\textbf{Panel (a)}: temperature dependence of the BLA for carbyne ($\Delta=0$ in the model), computed with the inclusion of QAE. We observe an hysteresis cycle which suggests that the polyyne-to-cumulene transition is of \textit{first order}, in contrast with the common Landau-Peierls picture of a second order transition, and consistently with the findings of Ref\cite{romanin2021dominant}. \textbf{Panel (b)}: free energy differences $\delta F=F_\mathrm{cooling}-F_\mathrm{heating}$ between the minimized free energies of the optimal structures obtained in the cooling cycle and those obtained in the heating cycle, shown in the range of temperatures where the hysteresis is found. From a linear fit of $\delta F$ with respect to $T$, we deduce a $T\mathrm{_{C}^{QAE}}\simeq4300\;\mathrm{K}$ above which the undimerized structure with no BLA becomes favorable.}
    \label{fig:BLA_vs_T}
\end{figure}

Next, we benchmark with respect to quantum-anharmonic thermal fluctuations effects in carbyne.
In Ref.\cite{romanin2021dominant}, hybrid-functional DFT + SSCHA calculations suggested the persistence of the dimerized polyyne phase up to very high temperatures, with an extrapolated lower bound for the critical temperature of $T_\mathrm{C}\sim3300\;\mathrm{K}$. The analysis of the temperature evolution of the quantum anharmonic free energy unveiled the coexistence of two minima in a broad temperature range, 
in stark contrast with the commonly accepted Landau-Peierls picture\cite{landau1937theorie,peierls1930theorie,landau2013course} of a second-order phase transition and suggesting instead a transition of the first order\cite{romanin2021dominant}.
Here, we investigate the temperature evolution of the BLA with QAE in the framework of the model. Starting from $T=0\;\mathrm{K}$, where the dimerized polyyne structure is the optimal configuration, we first performed an \textit{heating cycle} doing SSCHA minimizations for increasing values of the temperature. To test for the presence of meta-stable states, we used the optimal ionic density matrix obtained at a given temperature as the starting guess for the following one. As shown in Figure \ref{fig:BLA_vs_T}, at $T\simeq5000\;\mathrm{K}$ the cumulenic undimerized phase becomes eventually energetically favorable and the BLA disappears. Then, starting from a temperature where the cumulenic phase is stable, we performed a \textit{cooling cycle} down to $T=0\;\mathrm{K}$. Interestingly, it appears that cumulene is meta-stable down to $T\simeq600\;\mathrm{K}$.
Comparing the free energies of the optimal structures in the range of phase coexistence, we observe that polyyne is generally more favorable than cumulene, as shown in Figure \ref{fig:BLA_vs_T}. From a linear fit of the free energy difference $\delta F=F_\mathrm{cooling}-F_\mathrm{heating}$ with respect to the temperature $T$, we identify the value $T\mathrm{_{C}^{QAE}}\simeq4300\;\mathrm{K}$. The semi-quantitative agreement with the predicted lower bound $T_\mathrm{C}\gtrsim3300\;\mathrm{K}$ of Ref.\cite{romanin2021dominant} is a remarkable result, given the very simple form of the considered effective model.
These results put forward the model as a valid tool for investigating QAE on CPs, enabling fast \textit{and} reliable calculations and further validating its predictive power.

\section{Quantum-anharmonic effects on the structural properties} \label{sec2}

\begin{figure}[!htb]
    \begin{minipage}[c]{1.0\linewidth}
    \centering
    \includegraphics[width=\textwidth]{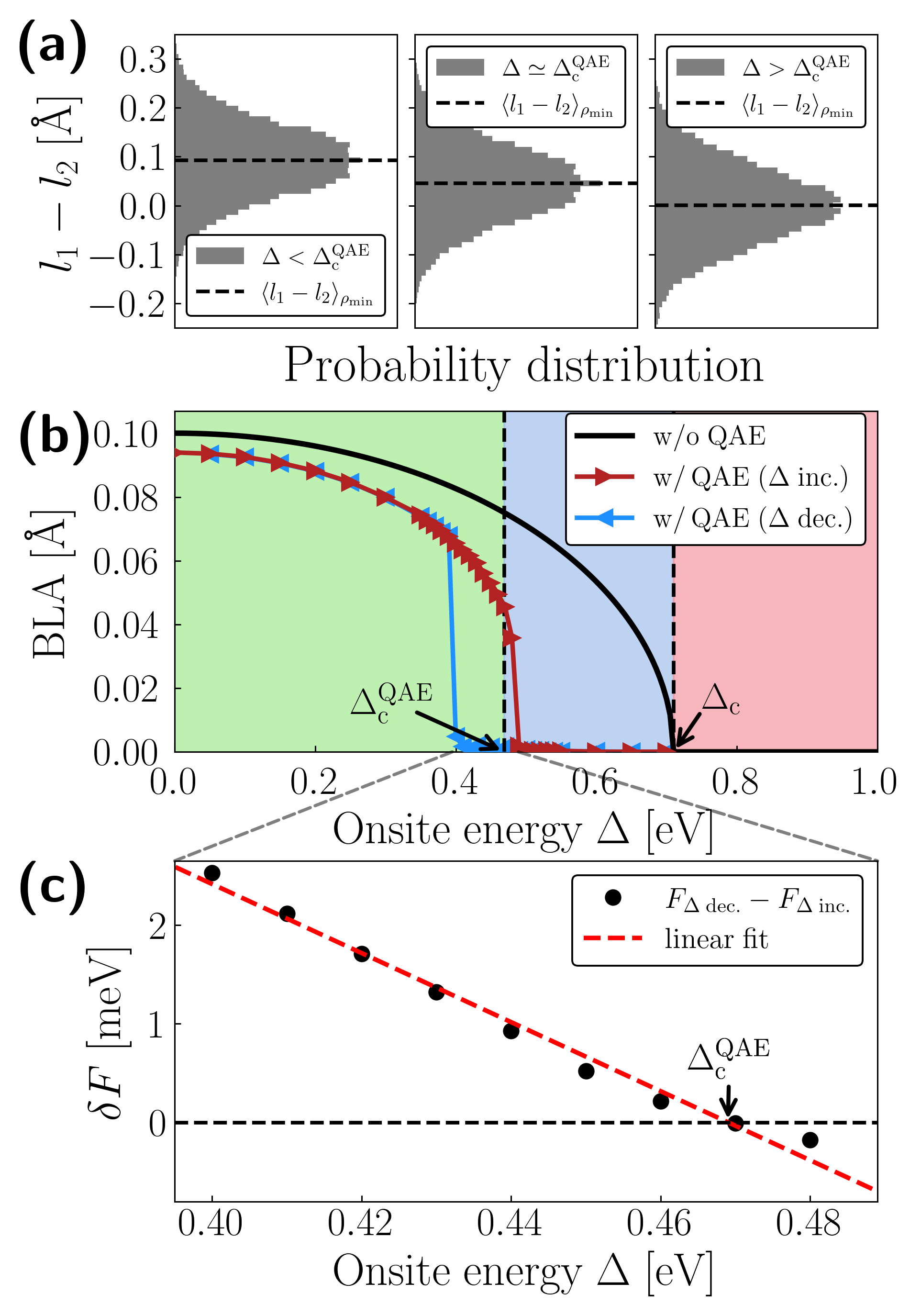}
    \end{minipage}
    \caption{
    Effects of quantum-anharmonicity on the manifestation of BLA in the model at $T=0\;\mathrm{K}$.
    \textbf{Panel (a)}:
    for three representative values of the onsite energy $\Delta=0\;\mathrm{eV}<\Delta_\mathrm{c}^\mathrm{QAE}$, $\Delta=0.47\;\mathrm{eV}\simeq\Delta_\mathrm{c}^\mathrm{QAE}$ and $\Delta=0.50\;\mathrm{eV}>\Delta_\mathrm{c}^\mathrm{QAE}$, we display the histograms with the values of the bond length differences $l_1-l_2$ computed along the supercell configurations as described in the text. 
    \textbf{Panel (b)}:
    effects of quantum anharmonicity on the displacive phase transition guided by the onsite energy $\Delta$. 
    The black line correspond to the values obtained neglecting the ionic fluctuations, whose effects are included in the values indicated by the triangles. To test for the presence of meta-stable states, we performed free energy minimizations for increasing and decreasing values of the onsite energy $\Delta$, as described in the text. We observe how the inclusion of QAE shifts the phase boundary between the lower-symmetric and the higher-symmetric phases. 
    \textbf{Panel (c)}:
    We obtain a value of $\mathrm{\Delta_{c}^{QAE}}\simeq0.47\,\mathrm{eV}$ from a linear fit of the free energy differences $\delta F=F_{\Delta\;\mathrm{dec.}}-F_{\Delta\;\mathrm{inc.}}$ between the free energy of the minimized configurations of the $\Delta$-decreasing cycle and those of the $\Delta$-increasing one, computed in the range of values of $\Delta$ where the two phases coexist.}
    \label{fig:BLA_vs_Delta}
\end{figure}

When the lattice dynamics is treated at a classical level, the optimal structural configuration at $T=0~\mathrm{K}$ is determined by the competition between two mechanisms that release the Peierls instability typical of these 1D systems by opening a gap in the electronic spectrum\cite{rice1982elementary,villani2024giant}. A finite e-ph coupling $\beta\neq0$ would favor the formation of symmetry-lowering BLA, whereas a finite onsite potential $\Delta\neq0$, accounting for inequivalent atoms along the chain, would keep the equally-spaced chain structure without BLA. As a consequence of this competition, increasing $\Delta$ at fixed e-ph coupling leads to a suppression of the BLA, resulting in a second-order phase transition at a critical value $\Delta_\mathrm{c}$ with BLA as the order parameter. The evolution of the order parameter is shown as a black line in Fig. \ref{fig:BLA_vs_Delta}(b), where we used model parameters listed in Table \ref{tab:fitted_parameters} and $\Delta$ as a variable. When QAE effects are included via the SSCHA, two major effects are observed. Firstly, BLA is significantly reduced, consistently with previous results on monoatomic chains\cite{Hirsch_prb1983_qf_ssh,Hanke_prb1986_qf_ssh,Barford_prb2006_qf_ssh,Bakrim_prb2007_qf_ssh}, and it vanishes at a lower value of $\Delta$ as compared to the classical result. Secondly, a region of coexistence between the two structural configurations develops, highlighted in Fig. \ref{fig:BLA_vs_Delta}(b) by the hysteresis cycle of BLA obtained for increasing (red line) and decreasing (blue line) values of $\Delta$. Comparison of the free energies of dimerized and undimerized phases in the coexistence region, shown in Fig. \ref{fig:BLA_vs_Delta}(c), unveils a first-order structural transition with a strongly renormalized critical value $\mathrm{\Delta_{c}^{QAE}}\simeq 0.66\times \Delta_c$.
The renormalization of BLA and the associated shift of the phase boundaries to a lower critical value for $\Delta$ arises from the quantum ionic fluctuations. 
To highlight their magnitude, 
in Figure \ref{fig:BLA_vs_Delta}(a) we display the $T=0$~K distributions of the bond length differences for representative values of the onsite energy: one is for monoatomic chains ($\Delta=0$), whereas the others are for diatomic chains in the proximity ($\Delta\simeq\Delta_\mathrm{c}^\mathrm{QAE}$) and slightly above ($\Delta=1.05\times\Delta_\mathrm{c}^\mathrm{QAE}$) 
the phase boundary, respectively. 
We observe how quantum-anharmonic fluctuations have a strong impact on the distribution of the atomic displacements. In fact, the values of the bond length differences fluctuate around the respective mean values with standard deviations $\sigma\simeq0.08$~\AA, comparable with the value of the $\mathrm{BLA}\simeq0.1$~\AA~obtained for carbyne in the absence of fluctuations (Table \ref{tab:fitted_parameters}).

\begin{figure}[!htb]
    \begin{minipage}[c]{1.0\linewidth}
    \centering
    \includegraphics[width=\textwidth]{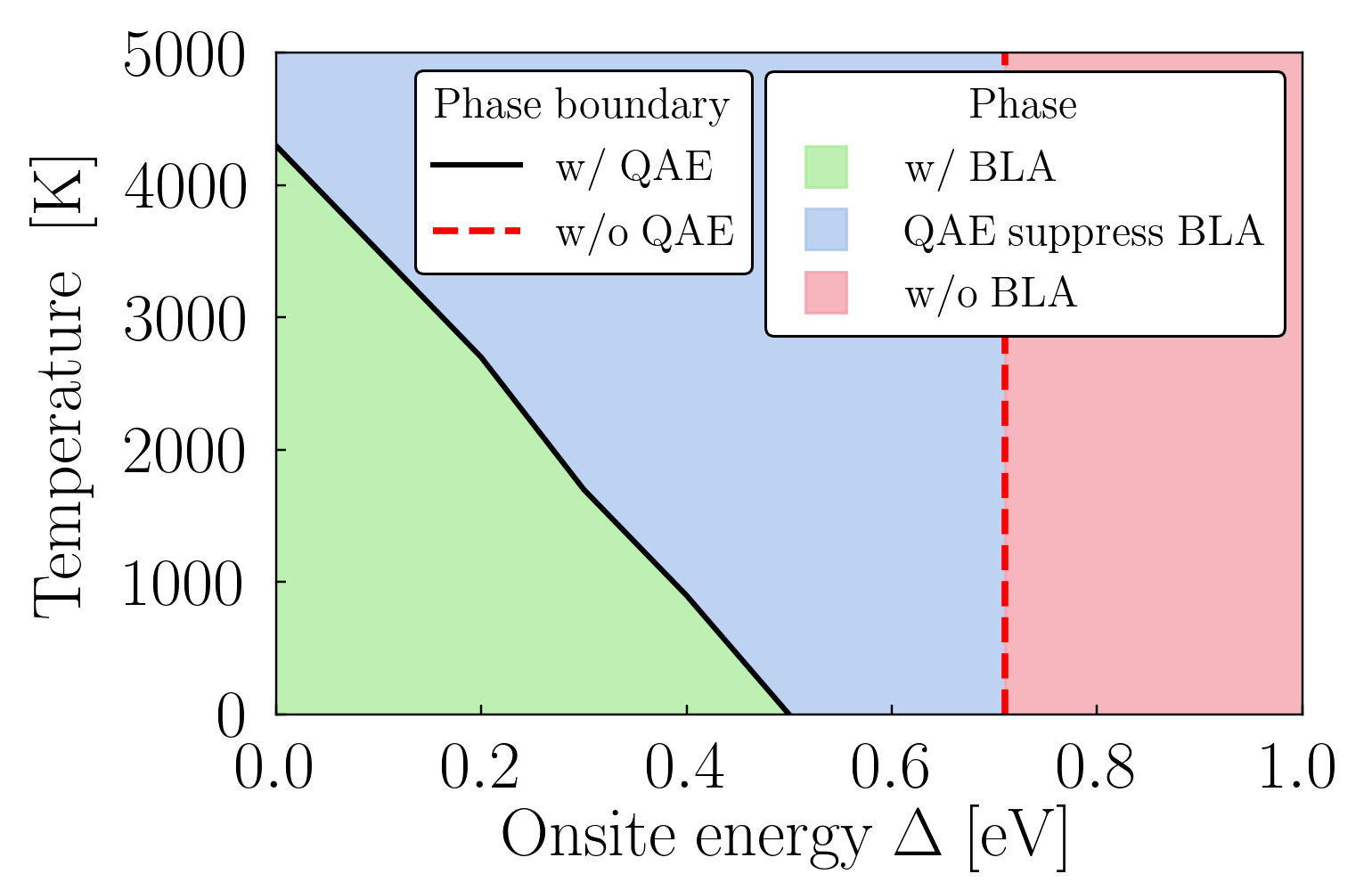}
    \end{minipage}
    \caption{Structural phase diagram of the model with respect to the temperature and the onsite energy. For different pairs of $T$ and $\Delta$, we display with different colors whether  a BLA is predicted or not. The green area represents the region for which the system manifests a BLA 
    and hence non-zero effective charges and a piezoelectric response, 
    whereas for values of $T$ and $\Delta$ in the pink region, the system presents no BLA and thus no piezoelectricity. 
    With light-blue we indicate the region where
    the inclusion of QAE suppresses the BLA. 
    The black line correspond to the values of $T\mathrm{_{C}^{QAE}}(\Delta)$ obtained with the inclusion of QAE as described in the text, representing the quantum-anharmonic-corrected phase boundary. The vertical dashed red line is in correspondence of the critical value of the onsite energy obtained at $T=0\;K$ without QAE.}
    \label{fig:phase_diagram}
\end{figure}

Finally, we expand on the results of Section \ref{sec1} on the temperature-dependence of the BLA in carbyne, as a representative of monoatomic chains, and further include the effects of finite temperatures on diatomic chains with modulated onsite potential. For different values of $\Delta$, we perform each time \textit{heating} and \textit{cooling} cycles as described in Section \ref{sec1}, obtaining different values of $T_\mathrm{C}(\Delta)$. In this way, we construct a boundary in the $\Delta$-$T$ space that separates the region where the diatomic chains display dimerization from the region without BLA.
The resulting structural phase diagram is shown in Fig. \ref{fig:phase_diagram}, where we also highlight in light blue the region where quantum-anharmonic effects suppress the BLA at zero temperature.
Even if the phase diagram has been obtained using parameters fitted on carbyne as a prototypical CP chain,
we expect that the uncovered phenomenology has a broader validity and that the framework we introduced can be applied also for the design of other functional CPs, such as PA or SPA.

\section{Quantum-anharmonic effects on the effective charges} \label{sec3}
\begin{figure}[!htb]
    \begin{minipage}[c]{1.0\linewidth}
    \centering
    \includegraphics[width=\textwidth]{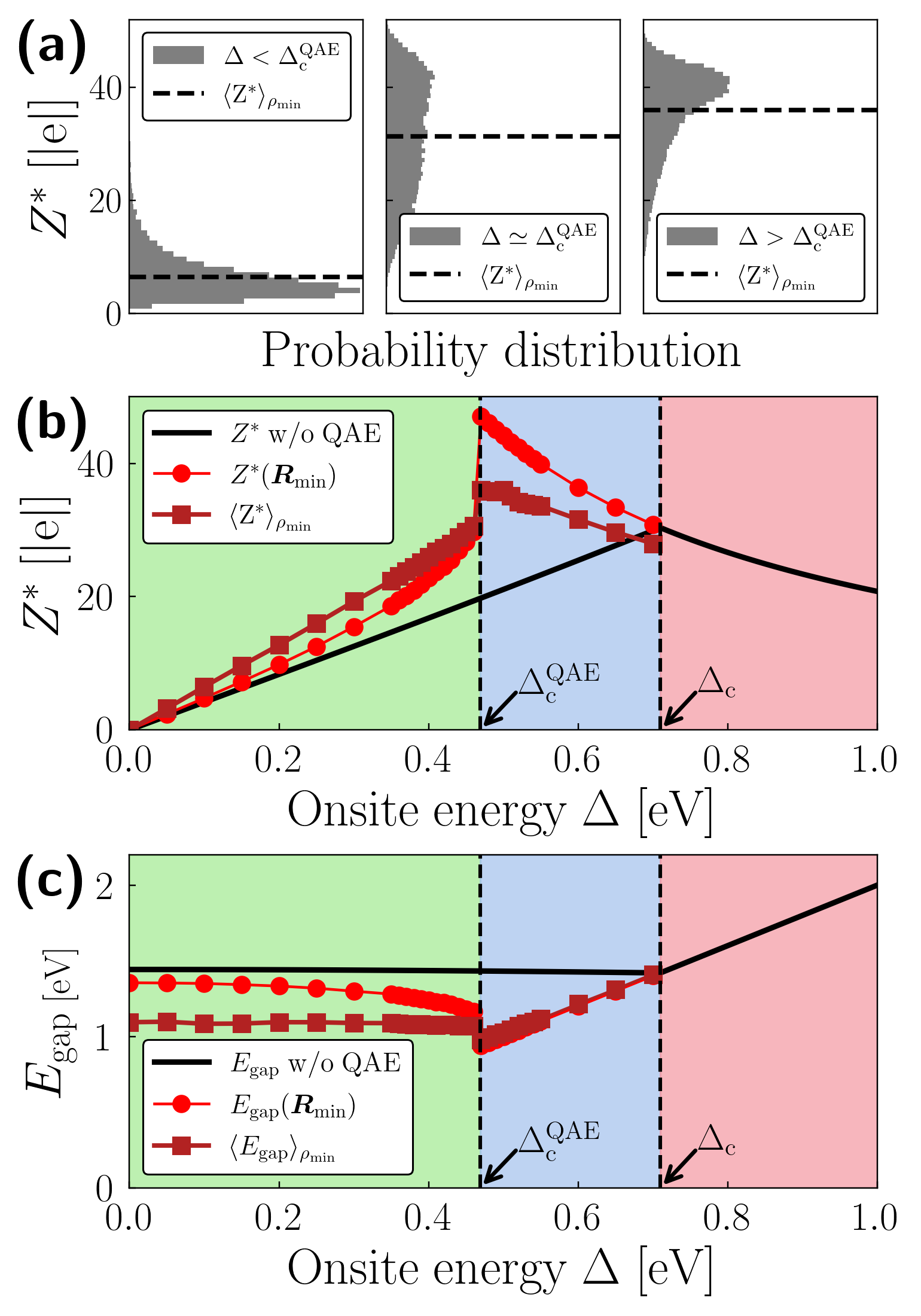}
    \end{minipage}
    \caption{
    \textbf{Panel (a)}: 
    for three representative values of onsite energies $\Delta=0.10\;\mathrm{eV}<\Delta_\mathrm{c}^\mathrm{QAE}$, $\Delta=0.47\;\mathrm{eV}\simeq\Delta_\mathrm{c}^\mathrm{QAE}$ and $\Delta=0.50\;\mathrm{eV}>\Delta_\mathrm{c}^\mathrm{QAE}$, we display the histograms with the values of the Born effective charges computed for all the supercell configurations extracted from the minimized density distributions $\rho_\mathrm{min}$. Ionic fluctuations result in the values of $Z^*$ varying over a large range, with a distribution that is broader in proximity of the phase boundary ($\Delta \simeq \Delta_\mathrm{c}^\mathrm{QAE}$). 
    \textbf{Panel (b)}: 
    behavior of the effective charges $Z^*$ as a function of the onsite energy difference $\Delta$, obtained with (marked lines) and without (black line) the inclusion of QAE, at $T=0\;\mathrm{K}$. The QAE-corrected values $\langle Z^* \rangle_{\rho_\mathrm{min}}$ and $ Z^*(\boldsymbol{R}_\mathrm{min})$ are, respectively, the averages over supercells configuration --Equation (\ref{eq:Zeff_QAE})-- and the effective charges of the system with atoms in the optimal configuration $\boldsymbol{R}_\mathrm{min}$. 
    \textbf{Panel (c)}: 
    same as the previous panel, for the values electronic energy gap $E_\mathrm{gap}$.
    Not only the topological behaviour of the effective charges --Equation (\ref{eq:Zeff_noQAE})-- is robust against the strong effects of ionic fluctuations, but their values attain a further enhancement --up to $\sim20\%$-- thanks to the reduction of the $E_\mathrm{gap}$ caused by QAE. 
    }
    \label{fig:Zeff_vs_D_QAE}
\end{figure}

We now turn our attention to the quantum-anharmonic effects on the Born effective charges, which quantify both the infrared response and the internal-relaxation contribution to the piezoelectric response.
The effective charge $Z^*_i$ of the atom $i$ along the chain is defined in Eq. (\ref{eq:zeff}) 
as the derivative of the electronic dipole moment per unit cell length $P$ with respect to the atomic displacement.
Because of the charge-neutrality acoustic sum rule
the chain must display electronically inequivalent atoms ($\Delta\neq0$) within its unit cell to have non-zero effective charges.
In the absence of fluctuations, the two atoms in each unit cell have effective charges of equal magnitude and opposite signs, allowing us to write $Z^*_i=(-1)^iZ^*$. The value of $Z^*$ can then be computed for any given atom, e.g., with linear perturbation theory, as shown in Appendix \ref{app_Zeff}.
In CPs, the high responsiveness and mobility of the delocalized electrons result in many distinctive features, and one of the most remarkable is arguably that the electronic charge density constitutes an exemplary case of a Thouless electronic pump\cite{thouless1983quantization,villani2024giant}. Recently\cite{villani2024giant}, we showed how this peculiar feature results in giant effective charges inversely proportional to the electronic band gap energy $E_\mathrm{gap}$, namely:
\begin{equation}
    \label{eq:Zeff_noQAE}
    Z^* \propto \beta\frac{\Delta}{E_\mathrm{gap}^2}, 
\end{equation}
with $E_\mathrm{gap}=2\sqrt{\Delta^2+\beta^2(l_1-l_2)^2}$. In Figure \ref{fig:Zeff_vs_D_QAE}(b) and \ref{fig:Zeff_vs_D_QAE}(c) we show with black lines the values of $Z^*$ and $E_\mathrm{gap}$ computed with respect to different onsite energy $\Delta$ in the absence of fluctuations. 
Since the order parameter evolves as $|l_1-l_2|\propto(\Delta-\Delta_\mathrm{c})^{1/2}$, it can be shown\cite{villani2024giant} that the energy gap is constant for values of $\Delta$ in the dimerized phase with bond-length alternation, and hence $Z^*\propto\beta\Delta$ --as also shown in Figure \ref{fig:fit_Zeff_piezo}-- whereas in the symmetric phase where $l_1=l_2$, it holds $Z^*\propto\beta/E_\mathrm{gap}=\beta/\Delta$. 

Thanks to the high responsiveness of the delocalized electrons, the effective charges attain giant values up to $30|e|$ when using parameters given in Table \ref{tab:fitted_parameters}, much larger than the expected effective charge of carbon of the order of the electronic charge $|e|$. As shown by the black line in Figure \ref{fig:Zeff_vs_D_QAE}(b), the maximum value for $Z^*$ is achieved at the phase boundary $\Delta_\mathrm{c}$.
However, this result relies on the assumption that the delocalized electrons are perturbed by a small and coherent structural modification, whereas 
strong fluctuations of the ionic positions represent a potentially detrimental factor. As discussed in the previous Section \ref{sec2}, ionic fluctuations are strong in the chain and the inclusion of QAE not only result in a shift of the phase boundary, but they also affect the order of the phase transition. For this reason, we study the impact of QAE on the effective charges in the model. If we consider the presence of ionic fluctuations, it is no longer true that $Z^*_i=(-1)^iZ^*$. However, exploiting the fact that (i) we can always distinguish two sublattices
thanks to their onsite energy $(-1)^i\Delta$, which is not affected by ionic fluctuations, and (ii) the sum rule $\sum_i Z^*_i = 0$ still holds when summing over all atoms in the supercell, we define the quantum-anharmonic corrected effective charge $Z^*$ as follows:
\begin{align}
    \label{eq:Zeff_QAE}
    \langle Z^* \rangle_{\rho} = \frac{1}{N_\mathrm{cells}N_\mathrm{conf}}\sum_{\mathcal{I}=1}^{N_\mathrm{conf}}\sum_{n=1}^{N_\mathrm{cells}}  {Z}^*_{2n}(\boldsymbol{R}_\mathcal{I})
\end{align}
where ${Z}^*_{2n}(\boldsymbol{R}_\mathcal{I})$ is the effective charge of atoms belonging to the sublattice indexed by $2n$ computed with linear response theory on the $\mathcal{I}$-th 
configuration $\boldsymbol{R}_\mathcal{I}$. We explicitly verified that 
if we substitute ${Z}^*_{2n}$ with ${Z}^*_{2n+1}$ we obtain the same result with just the opposite sign, thus fulfilling the charge-neutrality sum rule. 

In Figure \ref{fig:Zeff_vs_D_QAE}(b), we show as dark-red squares the values of $\langle Z^* \rangle_{\rho_\mathrm{min}}$ computed with the optimized ionic density distributions obtained from the free energy minimizations for different values of $\Delta$. 
Since thermal fluctuations become relevant at $T \gg T_\mathrm{room}$, as shown in Figure \ref{fig:BLA_vs_T}, here we consider only the $T=0$~K case.
In the region of phase coexistence, for each value of $\Delta$ we considered the configurations with the lowest minimized free energy. We observe that 
the effective charges still attain giant values that reach their maximum at the phase boundary, despite the latter being strongly shifted by QAE. Since the phase transition becomes of the first order when ionic fluctuations are included, the evolution of $Z^*$ displays a discontinuity at the phase boundary. When approaching the critical point from below, i.e., from the dimerized lower-symmetric phase, the effective charges are enhanced by QAE in such a way that they still approach the maximum value of $Z^*\simeq30|e|$ at the shifted phase boundary $\mathrm{\Delta_{c}^{QAE}}$. A similar enhancement is observed when approaching the critical point from above, i.e., from the undimerized phase, with a further enhancement up to $\sim20\%$ of the effective charge at the critical point. 
We argue that this effect is due to the topological character of the enhancement mechanism for the effective charge, which in the absence of fluctuations results in the inverse relation between $Z^*$ and $E_\mathrm{gap}$ of Eq. (\ref{eq:Zeff_noQAE}).
Indeed, ionic fluctuations are found to reduce the electronic energy gap $\langle E_\mathrm{gap} \rangle_{\rho_\mathrm{min}}$,
evaluated using Eq. (\ref{eq:QAE_average}) and shown in Figure \ref{fig:Zeff_vs_D_QAE}(c).

As discussed for the bond-length differences, the impact of ionic fluctuations on the effective charges can be appreciated from the distribution of their values, that we display in Figure \ref{fig:Zeff_vs_D_QAE}(a) for the same representative values of the onsite potential used in Figure \ref{fig:BLA_vs_Delta}. 
A peaked albeit asymmetric distribution is found away from the phase boundary, while strong QAE effects result in a very broad distribution near the critical point, again with a standard deviation of the same order of the average value of $Z^*$.
Another measure of the effects of ionic fluctuations 
is obtained by comparing the values of $\langle Z^* \rangle_{\rho_\mathrm{min}}$ with the values of the effective charge $Z^*(\boldsymbol{R}_\mathrm{min})$ computed on a chain with atoms in the optimal configuration $\boldsymbol{R}_\mathrm{min}=(\langle r_1 \rangle_{\rho_\mathrm{min}}, \dots, \langle r_{N_\mathrm{at}} \rangle_{\rho_\mathrm{min}})$, obtained from the SSCHA minimization, and shown as light-red circles in Figure \ref{fig:Zeff_vs_D_QAE}(b). In this definition, QAE enter in the response of the system only through the renormalization of the atomic positions. The difference between these two definitions can be appreciated in the context of linear-response theory within the SSCHA framework, where the quantum-anharmonic-corrected term $\langle Z^* \rangle_{\rho_\mathrm{min}}$ is the quantity that defines the infrared response of the system 
including one-phonon scattering processes, which are instead neglected in $Z^*(\boldsymbol{R}_\mathrm{min})$\cite{monacelli2021time,siciliano2023wigner}. The comparison shown in Figure \ref{fig:Zeff_vs_D_QAE}(b) indeed confirms non-negligible renormalization effects due to QAE. Interestingly, the relation between effective charge and energy gap of Eq. (\ref{eq:Zeff_noQAE}) still holds qualitatively between $Z^*(\boldsymbol{R}_\mathrm{min})$ and $E_\mathrm{gap}(\boldsymbol{R}_\mathrm{min})$, which is also shown as light-red circles in Figure \ref{fig:Zeff_vs_D_QAE}(c), further supporting the topological protection of the enhancement mechanism for $Z^*$. 
We conclude this section by noticing that the quantity $\langle \sum_iZ^*_ir_i \rangle_{\rho_\mathrm{min}}$ can be used to assess the effect of
two-phonon scattering processes to the infrared response\cite{monacelli2021time,siciliano2023wigner}.
In Appendix \ref{app_piezo}, we show that $\langle \sum_iZ^*_ir_i \rangle_{\rho_\mathrm{min}} \simeq \sum_i\langle Z^*_i \rangle_{\rho_\mathrm{min}} \langle r_i \rangle_{\rho_\mathrm{min}}$, that we compute for different values of the onsite energy $\Delta$. Our results, shown in Figure \ref{fig:DeltaP_vs_D}, suggest that the two-phonon effects to the infrared response are negligible in the model. 

\section{Quantum-anharmonic effects on the piezoelectric response} \label{sec4} 
In this Section we finally discuss the effects of quantum anharmonicity on the morphotropic-like and topological contributions to the enhancement mechanism of CPs piezoelectric response.
As introduced in Section \ref{sec2}, the effect of a strain $\epsilon$ on the electronic charge density of the chain is quantified through the piezoelectric coefficient $c_\mathrm{piezo}$, defined in Equation (\ref{eq:def_piezo}) as the total derivative of the polarization $P$ with respect to $\epsilon$. 
When subject to a uniform strain, the elementary cells of the 1D chain can only present contractions/dilatations, described by the relation $a(\epsilon)=a_0(1+\epsilon)$, where $a_0$ indicates the length of the cell in the absence of strain ($\epsilon=0$). 
Following the standard approach\cite{piezoMartin1972,wu2005systematic}, we decompose the response of the system to a strain into a contribution where the atoms in the cells move rigidly with the deformation --the so-called \textit{clamped ion} term $c_\mathrm{piezo}^\mathrm{c.i.}$-- and another contribution which accounts for the fact that the ions relax and occupy different positions inside the strained cells, quantified by the \textit{internal-relaxation} term $c_\mathrm{piezo}^\mathrm{i.r.}$. 
In terms of the internal fractional coordinates $\boldsymbol{u}=\boldsymbol{R}/a$ and exploiting the fact that the polarization $P(\epsilon,\boldsymbol{u}(\epsilon))$ depends on the strain both explicitly and implicitly through $\boldsymbol{u}(\epsilon)$, the total derivative of Equation (\ref{eq:def_piezo}) can be decomposed as follows:
\begin{align}
    c_\mathrm{piezo} &= c_\mathrm{piezo}^\mathrm{c.i.} + c_\mathrm{piezo}^\mathrm{i.r.}, \\
    c_\mathrm{piezo}^\mathrm{c.i.} &= \frac{\partial P}{\partial \epsilon} \label{eq:def_clam_pion}, \\ 
    c_\mathrm{piezo}^\mathrm{i.r.} &= \sum_i\frac{\partial P}{\partial u_i}\frac{\partial u_i}{\partial \epsilon} \\
    &=\sum_i Z_i^*\frac{\partial u_i}{\partial \epsilon} \label{eq:def_int_rel}.
\end{align}
The last equality, obtained via Equation (\ref{eq:zeff}) and the definition of internal coordinates, shows that the internal-relaxation contribution stems from the composition of Born effective charges and internal strain ${\partial u_i}/{\partial \epsilon}$.
In the absence of fluctuations, the internal-relaxation contribution in the diatomic chain may attain arbitrarily large values thanks to the topological enhancement of the effective charges combined with the diverging behaviour of the internal-strain when approaching the second-order structural phase-transition point\cite{villani2024giant}.

While the topological enhancement of the effective charges is stable against quantum anharmonic fluctuations, as discussed in Section \ref{sec3}, the structural phase transition is instead sensitive to QAE, as seen in Sections \ref{sec1}E and \ref{sec2},
possibly affecting the diverging behavior of the internal-strain derivative and consequently threatening the morphotropic-like enhancement of the internal-relaxation piezoelectric response. To shed light on this issue and estimate the effects of ionic fluctuations on this enhancement mechanism, we compute the complete piezoelectric coefficient as well as the internal-relaxation contribution, with and without the inclusion of QAE. In the absence of fluctuations, $c_\mathrm{piezo}$ is calculated as finite differences of $P$, namely
\begin{equation}
    c_\mathrm{piezo} = \frac{ P (+\epsilon) - P(-\epsilon)}{2 \epsilon},
\end{equation}
where $P(\pm\epsilon)$ is the dipole moment per cell length of the elementary diatomic cell of the chain under an applied homogeneous strain of $\pm\epsilon$, respectively, that we compute using the Berry phase approach\cite{berry1984quantal,resta1994macroscopic}.  
Following the same procedure described in the previous Section, we first account for the effects of ionic fluctuations through the renormalization of the atomic positions only, by computing the piezoelectric coefficient 
\begin{align}
    \label{eq:cpiezo_rmin}
    c_\mathrm{piezo}(\boldsymbol{R}_\mathrm{min}) = \frac{ P(\boldsymbol{R}_\mathrm{min}(+\epsilon)) - P(\boldsymbol{R}_\mathrm{min}(-\epsilon))}{2 \epsilon}
\end{align}
where $\boldsymbol{R}_\mathrm{min}(\pm\epsilon)$ are the atomic positions obtained from the
the minimization of the free energy of a system with an applied strain of $\pm\epsilon$. 
Then, we compute the piezoelectric coefficient $\langle c_\mathrm{piezo} \rangle_{\rho_\mathrm{min}}$ averaged over the ionic quantum fluctuations as
\begin{align}
    \label{eq:cpiezo_avg}
    \langle c_\mathrm{piezo} \rangle_{\rho_\mathrm{min}} = \frac{\langle P \rangle_{\rho_\mathrm{min}(+\epsilon)} - \langle P \rangle_{\rho_\mathrm{min}(-\epsilon)}}{2 \epsilon}
\end{align}
where $\langle P \rangle_{\rho_\mathrm{min}(\pm\epsilon)}$ is the quantum-anharmonic corrected value of $P$ computed over the optimal density distribution $\rho_\mathrm{min}(\pm\epsilon)$. Finally, we also compute $\langle c_\mathrm{piezo}^\mathrm{i.r.} \rangle_{\rho_\mathrm{min}}=\langle \sum_i Z_i^* (\partial u_i / \partial\epsilon) \rangle_{\rho_\mathrm{min}}$, as detailed in Appendix \ref{app_piezo}.

\begin{figure}[!ht]
    \begin{minipage}[c]{1.0\linewidth}
	\centering
    \includegraphics[width=1.0\textwidth]{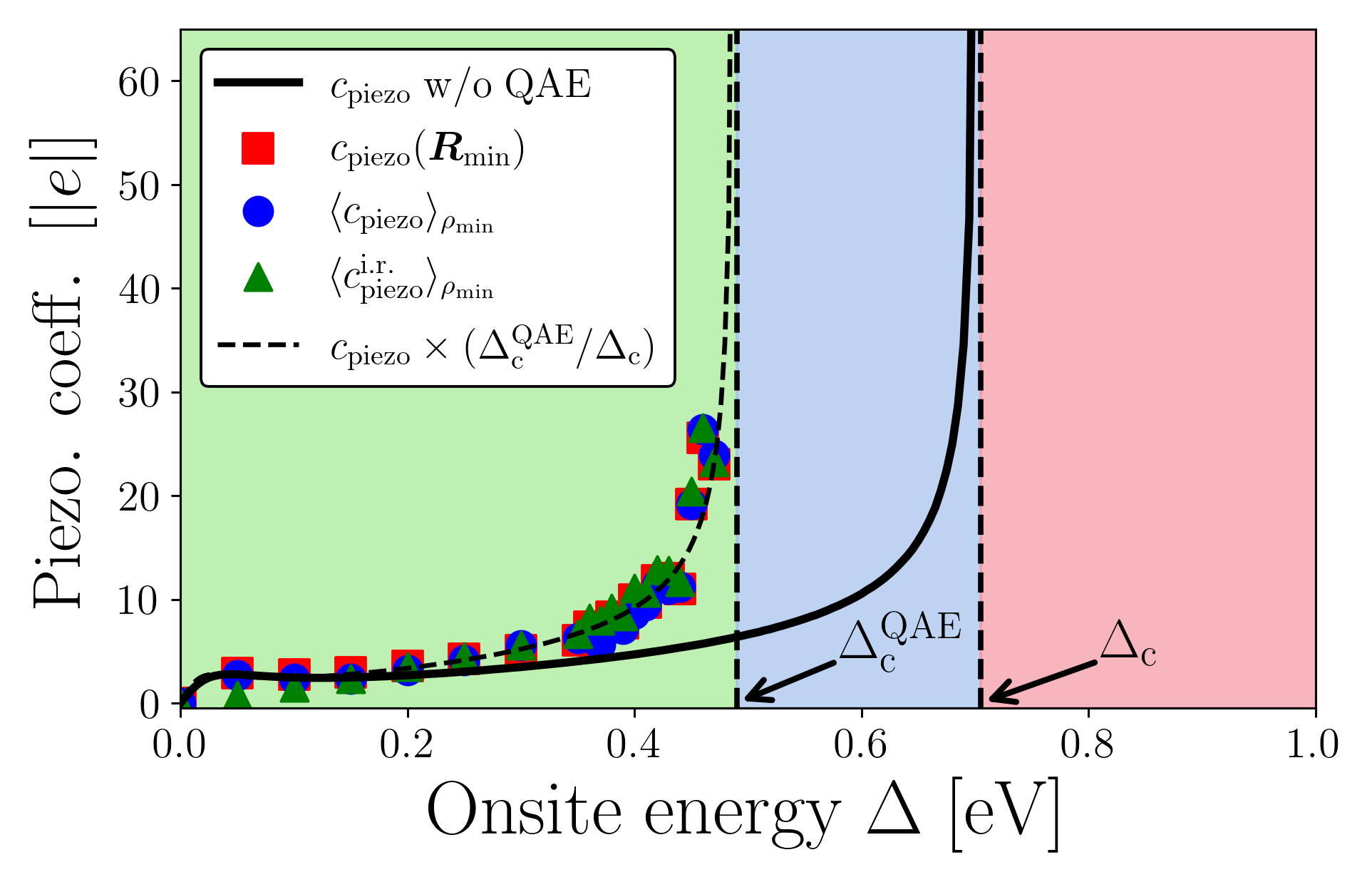} 
    \caption{
    Evolution of the piezoelectric coefficient as a function of the onsite energy $\Delta$, with and without QAE (coloured dots and solid black line, respectively).
    The red squares and the blue dots represent, respectively, the total piezoelectric response
    $c_\mathrm{piezo}$ computed on the structures obtained from the free energy minimizations --Eq. (\ref{eq:cpiezo_rmin})-- and the values computed as averages on the supercell configurations --Eq. (\ref{eq:cpiezo_avg})--, as explained in the text. The green triangles are the values of the internal-relaxation contribution $\langle c_\mathrm{piezo}^\mathrm{i.r.} \rangle_{\rho_\mathrm{min}}$, which dominates the total piezoelectric response. The dashed line corresponds to the behaviour of $c_\mathrm{piezo}$ computed in the absence of fluctuations rescaled for the ratio $\Delta_\mathrm{c}^\mathrm{QAE}/\Delta_\mathrm{c}$, highlighting how the morphotropic-like behaviour of the piezoelectric response is robust against QAE.
    }
	\label{fig:piezo_comparison_QAE} 
    \end{minipage} 
\end{figure} 

In Figure \ref{fig:piezo_comparison_QAE} we compare the values of $c_\mathrm{piezo}$, $c_\mathrm{piezo}(\boldsymbol{R}_\mathrm{min})$, $\langle c_\mathrm{piezo} \rangle_{\rho_\mathrm{min}}$ and $\langle c_\mathrm{piezo}^\mathrm{i.r.} \rangle_{\rho_\mathrm{min}}$, obtained for different values of the onsite energy $\Delta$. As for the effective charges, we considered the $T=0$~K case only
and in the region of phase coexistence we considered the configurations with the lowest minimized free energy. 
As expected, the fact that QAE reduce the order of the transition is reflected by the absence of a truly diverging behaviour when approaching the quantum-anharmonic-corrected critical value. Nevertheless, we still observe a giant enhancement of the piezoelectric coefficient in the proximity of the phase boundary, suggesting that the morphotropic-like character of the piezoelectric response of the chain remains robust against quantum-anharmonicity in a large region of onsite energy values. Indeed, the increase of the piezoelectric coefficient while approaching the transition point from below, i.e., from the dimerized phase, is still captured by the same functional form obtained in the absence of quantum lattice fluctuations upon a simple rescaling with the ratio $\Delta_\mathrm{c}^\mathrm{QAE}/\Delta_\mathrm{c}$, shown as a dashed line in Figure \ref{fig:piezo_comparison_QAE}. On the other hand, the effect of quantum anharmonic fluctuations on the total piezoelectric response beyond the renormalization of the atomic positions, which can be quantified -- as in Section \ref{sec3} for effective charges -- by comparing $c_\mathrm{piezo}(\boldsymbol{R}_\mathrm{min})$ and $\langle c_\mathrm{piezo} \rangle_{\rho_\mathrm{min}}$, is essentially negligible. As a further confirmation of the morphotropic-like enhancement, we remark that
the piezoelectric response is still dominated by the internal-relaxation contribution, displayed as green triangles in Figure \ref{fig:piezo_comparison_QAE}. 

\section{Conclusions} \label{sec5} 
In the present work, we analyzed the impact of ionic quantum-anharmonic effects on the giant longitudinal piezoelectric response recently predicted in functionalized conjugated polymers. To this end, we examined how ionic fluctuations influence several intertwined aspects of conjugated-chain physics. We combined the stochastic self-consistent harmonic approximation with a Rice–Mele diatomic-chain model parametrized to reproduce hybrid-functional first-principles calculations of carbyne. The resulting framework proves both efficient and predictive, yielding semi-quantitative agreement with fully first-principles calculations of polar responses in decorated carbyne, as well as with literature results on quantum-anharmonic and thermal effects in carbyne. Within this approach, we find that quantum anharmonicity has a pronounced impact on structural properties, with bond-length differences that fluctuate on the same scale of the average BLA. Moreover, the critical point for the composition-driven dimerization structural transition is displaced by more than $30\%$, with metastable solutions appearing in a broad region in the phase space, signalling the first-order character of the phase transition.

Despite this strong structural renormalization, the mechanisms responsible for the large piezoelectric response remain robust against QAE. The persistence of giant Born effective charges--enhanced near the renormalized critical region and correlated with a fluctuation-induced reduction of the electronic gap--indicates that the Thouless-pump character of the $\pi$-electrons polarization is not only protected against ionic fluctuations but it is even reinforced. In parallel, the morphotropic-like behavior of the piezoelectric response survives. The piezoelectric coefficient retains a pronounced maximum in the vicinity of the renormalized boundary and remains dominated by the internal-relaxation contribution, with quantum anharmonicity primarily shifting and reshaping the optimal enhancement window rather than suppressing it. 

Overall, our results further establish quantum-anharmonic effects as essential for a reliable \textit{in silico} description of conjugated systems, in particular for their structural properties. At the same time, the robustness of the enhancement mechanisms supports functionalized conjugated polymers as a viable platform for highly responsive organic materials for electromechanical applications.

\section*{Acknowledgements} 
The authors acknowledge financial support from the European Union under ERC-SYN MORE-TEM, No. 951215, and from the Italian MIUR through PRIN-2017 project, Grant No. 2017Z8TS5B.
We also acknowledge CINECA awards under ISCRA initiative Grant No. HP10CCJFWR and HP10C7XPLJ for the availability of high performance computing resources and support. Views and opinions expressed are however those of the author(s) only and do not necessarily reflect those of the European Union or the European Research Council. Neither the European Union nor the granting authority can be held responsible for them.

\section*{Competing interests}
The Authors declare no Competing Financial or Non-Financial Interests.

\section*{Data availability}
The original data for each figure are available from the corresponding author upon reasonable request.

\bibliographystyle{unsrt} 
\bibliography{bibliography} 

\appendix 
\section{Details on the calculation of forces and energies in the model} \label{app:model_forces_energies} 
In this Appendix, we show how to compute the electronic energies and forces used for the SSCHA minimization procedure. We consider a supercell with $N_\mathrm{cells}$ elementary diatomic cells and $N_\mathrm{at} = 2 \times N_\mathrm{cells}$ atoms, constructed as described in the main text. For a given atomic configuration $\boldsymbol{R}=(r_1,\dots,r_{N_\mathrm{at}})$ and model's parameters $t_0$, $\beta$, $\mathrm{K}$, and $\Delta$, we compute the total energy $E_\mathrm{tot}(\boldsymbol{R};t_0,\beta,\mathrm{K},\Delta)$ of the supercell from the Hamiltonian defined in Equation (\ref{eq:H_tot}), following the SSCHA approach\cite{monacelli2021stochastic}. Implying the dependence on all the parameters, that we avoid writing explicitly for clarity, we separate $E_\mathrm{tot} = E_\mathrm{L} + E_\mathrm{e}$ into a term $E_\mathrm{L}$ that accounts for the lattice distortion and a term $E_\mathrm{e}$ that accounts for the contribution of the $\pi$-electrons in the tight-binding model.We compute the former as 
\begin{equation}
    E_\mathrm{L} = \sum_{i=1}^{N_\mathrm{at}} \frac{1}{2}\mathrm{K} \delta r_{i+1,i} ^2
\end{equation}
whereas we obtain the latter from the Hamiltonian $H_\mathrm{e}$, which reads
\begin{align}
    H_\mathrm{e} &= 2n_\mathrm{e} \Delta(-1)^ic^\dagger_ic^{\phantom\dagger}_i + \nonumber \\ & -2n_\mathrm{e} \left[\left( t_0 -\beta\delta r_{i+1,i}  \right) c^\dagger_{i+1} c^{\phantom\dagger}_i + \mathrm{h.c.} \right].
\end{align} 
In particular, following the standard procedure, we transform $H_\mathrm{e}$ from the real to the reciprocal space considering $N_k$ $k$-points defined over the first Brillouin zone of the supercell. In this way, we obtain $N_k$ $k$-dependent $N_\mathrm{at} \times N_\mathrm{at}$ Hamiltonian matrices $H_{\mathrm{e},k}$, namely
\begin{align}
    \label{eq:H_elec}
    H_{\mathrm{e},k} =
    \begin{bmatrix}
    -\Delta   & T_1       & 0       & \dots  & 0           & T^{*}_{N_\mathrm{at}} \\ 
    T_{1}^{*} & +\Delta   & T_{2}   & \ddots & \ddots      & 0         \\
    0         & T_{2}^{*} & -\Delta & \ddots & \ddots      & \vdots    \\
    \vdots    & \ddots    & \ddots  & \ddots & \ddots      & 0         \\
    0         & \ddots    & \ddots  & \ddots & -\Delta     & T_{N_\mathrm{at}-1}   \\
    T_{N_\mathrm{at}}     & 0       & \dots  & 0           & T_{N_\mathrm{at}-1}^{*} & +\Delta   \\
    \end{bmatrix},
\end{align}
where
\begin{align} 
    \label{eq:def_T_alpha}
    T_{i+1,i} 
    &= -t_{i+1,i}\e^{\iu k (r_{i+1} - r_{i})} \\ 
    &=
    [-t_0 + \beta \delta r_{i+1,i} ]\e^{\iu k (\delta r_{i+1,i} + a/2)}.
\end{align}
We diagonalize for each $k$ in the first Brillouin zone of the supercell to obtain the $N_k \times N_\mathrm{at}$ eigenvalues $\varepsilon_{i,k}$. Then, we obtain the contribution $E_\mathrm{e}$ to the total energy summing over the $N_k\times N_\mathrm{occ}$ occupied states as
\begin{equation}
    \label{eq:E_e}
    E_\mathrm{e} = \frac{2 n_\mathrm{e}}{N_k} \sum_{k=1}^{N_k}\sum_{j=1}^{N_\mathrm{occ}} \varepsilon_{j,k},
\end{equation}
where the factor of $2$ is for the spin and $n_\mathrm{e}$ accounts for the number of $p$-orbitals per atoms, holding $n_\mathrm{e}=1$ for (S)PA, and $n_\mathrm{e}=2$ for (decorated) carbyne, and $N_\mathrm{occ}=N_\mathrm{at}/2$ is the number of occupied bands. We follow a similar procedure for the calculation of the forces. The total force $F_{\mathrm{tot},i}$ acting on atom $i$ is obtained from the derivative of the total energy with respect to the displacement of atom $i$, namely
\begin{align}
    F_{\mathrm{tot},i} = -\frac{\partial E_\mathrm{tot}}{\partial r_i}
\end{align}
and again we separate $F_{\mathrm{tot},i} = F_{\mathrm{L},i} + F_{\mathrm{e},i}$ into the lattice and the electronic contributions. 
The former reads 
\begin{equation}
    F_{\mathrm{L},i} = -\mathrm{K} ( 2r_i - r_{i+1} - r_{i-1} ),
\end{equation}
whereas the second is computed using the Hellman-Feynman theorem as
\begin{equation}
    F_{\mathrm{e},i} = - \frac{2 n_\mathrm{e}}{N_k} \sum_{k=1}^{N_k}\sum_{j=1}^{N_\mathrm{occ}} \langle \psi_{j,k}| \frac{\partial H_{\mathrm{e},k}}{\partial r_i} | \psi_{j,k}\rangle
\end{equation}
with $|\psi_{j,k}\rangle$ being the eigenvector associated to $\varepsilon_{j,k}$.

\section{Details on the construction of the model} \label{app:model_fit}
In the absence of fluctuations, we considered a single diatomic unit cell ($N_\mathrm{cells}=1$) in periodic boundary conditions with $N_k=320$. 
We obtained the values reported in Table \ref{tab:fitted_parameters} by minimizing the following loss function
\begin{equation}
    \label{eq:loss_function}
    \mathcal{L}\left(t_0,\beta,\mathrm{K}\right) = \sum_{i} \left[\frac{ Q_{i}\left(t_0,\beta,\mathrm{K}\right) - Q^\mathrm{DFT}_{i} }{ Q^{\mathrm{DFT}}_i} \right]^2,
\end{equation}
where $Q_i$ indicates the quantities that we considered, i.e. the BLA, the depth of the total energy profile $E_{\mathrm{gain}}$ and the frequency at $\Gamma$ of the longitudinal optical phonon of carbyne $\omega_{\mathrm{LO}}(\Gamma)$. In particular, $Q_{i}\left(\beta,\mathrm{K},t_0\right)$ is the value calculated within the model, whereas $Q^\mathrm{DFT}_{i}$ indicates the target DFT@PBE0 value. At each step of the minimization procedure of the loss function, quantities within the model where computed as follows: the BLA was obtained by minimizing the total energy $E_\mathrm{tot}$ with respect to the atomic positions and using the relation of Equation (\ref{eq:def_BLA_tm}). $E_{\mathrm{gain}}$ was computed as the difference between the total energy of the chain with equidistant atoms and the chain with the BLA obtained as just described. $\omega_{\mathrm{LO}}(\Gamma)$ was computed diagonalizing the dynamical matrix obtained from the force-constants matrix, i.e., the mass-weighted matrix of the second order derivatives of the total energy with respect to the atomic positions, using the atomic mass of the carbon atom. The value of the loss obtained with the parameters of Table \ref{tab:fitted_parameters} is $\mathcal{L}\left(t_0,\beta,\mathrm{K}\right)<10^{-8}$. 
We obtained the values of $\Delta$ reported in Table \ref{tab:fit_Delta} minimizing the following loss function 
\begin{equation}
    \label{eq:loss_function_D}
    \mathcal{L}\left(\Delta\right) = \sum_i \left[ \frac{ Q_{i}\left(\Delta\right) - Q^\mathrm{DFT}_{i} }{ Q^{\mathrm{DFT}}_i }\right]^2 
\end{equation}
where $Q_{i}$ indicates the quantities that we used, i.e. the BLA and the depth of the total energy profile $E_{\mathrm{gain}}$, $Q_{i}\left(\Delta\right)$ are the values computed within the model with the parameters $\beta$, $t_0$ and $\mathrm{K}$ of Table \ref{tab:fitted_parameters}, and $Q^\mathrm{DFT}_{i}$ are the target DFT@PBE0 values. In Table \ref{tab:fit_Delta_loss} we report the values of the loss function obtained for each fitting procedure.
\begin{table}[htb]
    \centering
    \begin{tabular}{ccc}
    \hline \hline
    \rule{0pt}{1.0em}
    \hspace{
    0.1cm} $d_{\mathrm{He-C}}$ (\AA) \hspace{0.1cm} &  \hspace{0.1cm} $\Delta$ (eV) \hspace{0.1cm} &  \hspace{0.1cm} Loss  \hspace{0.1cm} 
    \rule[-3.0pt]{0pt}{0pt} \\
    \hline\hline
    1.80 & 0.521 & 0.175 \\
    2.00 & 0.287 & 0.020 \\
    2.20 & 0.136 & 0.002 \\
    2.40 & 0.047 & 0.002 \\
    2.60 & 0.015 & 0.002 \\
    2.80 & 0.013 & 0.002 \\
    3.00 & 0.012 & 0.001 \\
    \hline \hline
    \end{tabular}
    \caption{For each value of the distance $d_{\mathrm{He-C}}$ between the Helium and Carbon atoms as defined in the main text, we fitted the values of $\Delta$ minimizing the loss function of Equation (\ref{eq:loss_function_D}). For smaller distances $d_{\mathrm{He-C}}$, the minimized loss is higher because the effects of the electronic orbitals of the Helium atoms on the parameters $t_0$, $\beta$, and $\mathrm{K}$ become non-negligible.}
    \label{tab:fit_Delta_loss}
\end{table}

\section{\textit{Ab initio} computational details} \label{sec:app_comp_detail}
All DFT@PBE0 calculations for carbyne and decorated carbyne were performed using the CRYSTAL code\cite{dovesi2014crystal14,dovesi2018quantum}. This code allows for the simulation of truly isolated systems, as the 1D CPs addressed in our work. We considered a unit cell with $2$ Carbon atoms for carbyne, and $2$ Carbon atoms and $6$ Helium atoms for decorated carbyne. We used triple-$\zeta$-polarised Gaussian-type bases\cite{vilela2019bsse} with real space integration tolerances of 10-10-10-15-30 and an energy tolerance of $10^{-10}$ Ha for the total energy convergence. In the most general case, the Born effective charges are a $3\times3\times N_\mathrm{atoms}$ tensor that within the CRYSTAL code are computed as finite differences of polarization.
In Figure \ref{fig:fit_Zeff_piezo} of the main text, we displayed the values $Z^*_{xx,\mathrm{C_\mathrm{He}}}$ of the $xx$ component, i.e. the component parallel to the chain direction, of the Born effective charge tensor of the Carbon atom surrounded by the Helium atoms. The \textit{ab initio} piezoelectric coefficients $c_{\mathrm{piezo}}$ of Figure \ref{fig:fit_Zeff_piezo} were computed using the Berry-phase approach\cite{vanderbilt2000berry} as implemented in the code\cite{erba2013piezoelectricity,erba2016internal}, which accounts also for transverse displacements. 

\section{Details on the calculation of the effective charges} \label{app_Zeff}
We computed the Born effective charges using linear perturbation theory as 
\begin{equation}
    \label{eq:def_Zeff_perturbation}
    Z^*_i = \frac{4n_\mathrm{e}}{N_k}\sum_k^{N_k}\sum_{l}^{N_\mathrm{occ}}\sum_{m}^{N_\mathrm{emp}} \frac{\braket{\psi_{l,k}|\frac{\partial H_{\mathrm{e},k}}{ \partial r_i} |\psi_{m,k}} \braket{\psi_{m,k}| \frac{H_{\mathrm{e},k}}{ \partial k} |\psi_{l,k}}}{(\varepsilon_{l,k}-\varepsilon_{m,k})^2 },
\end{equation}
where the factor $4$ comes from the spin degeneracy and from the sum over the complex conjugate of a real quantity, and the sums run over the $N_k$ $k$-points, the $N_\mathrm{occ}$ occupied bands, and the $N_\mathrm{emp}$ empty bands, respectively. For the results obtained in the absence of ionic fluctuations, we considered a single diatomic unit cell ($N_\mathrm{cells}=1$) in periodic boundary conditions with $N_k=320$. Thanks to the charge-neutrality sum rule, it holds $Z^*_1 = -Z^*_2$ for the atoms in the cell, implying that the calculation of just one of the two values is enough. In the presence of fluctuations, this latter equality doesn't hold, so we computed all the values $Z^*_i$ for the $N_\mathrm{at}$ atoms in the $N_\mathrm{cells}$ diatomic cells of the $N_\mathrm{conf}$ supercells, using Equation (\ref{eq:def_Zeff_perturbation}). The quantity $\langle Z^* \rangle_{\rho_\mathrm{min}}$ defined in the main text was computed using $N_\mathrm{conf}=504$. To obtain the values of $Z^*(\boldsymbol{R}_\mathrm{min})$, we considered a single diatomic unit cell ($N_\mathrm{cells}=1$) with the atoms in positions $\langle r_1 \rangle_{\rho_\mathrm{min}}$ and $\langle r_2 \rangle_{\rho_\mathrm{min}}$, respectively, since it holds $\langle r_{2n} \rangle_{\rho_\mathrm{min}}=\langle r_2 \rangle_{\rho_\mathrm{min}}$ and $\langle r_{2n+1} \rangle_{\rho_\mathrm{min}}=\langle r_1 \rangle_{\rho_\mathrm{min}}$ $\forall n=1,\dots,N_\mathrm{cells}$. 

\section{Details on the SSCHA applied to the model} \label{app_SSCHA_model} 
\begin{figure}[!ht]
    \begin{minipage}[c]{1.0\linewidth}
	\centering
    \includegraphics[width=1.0\textwidth]{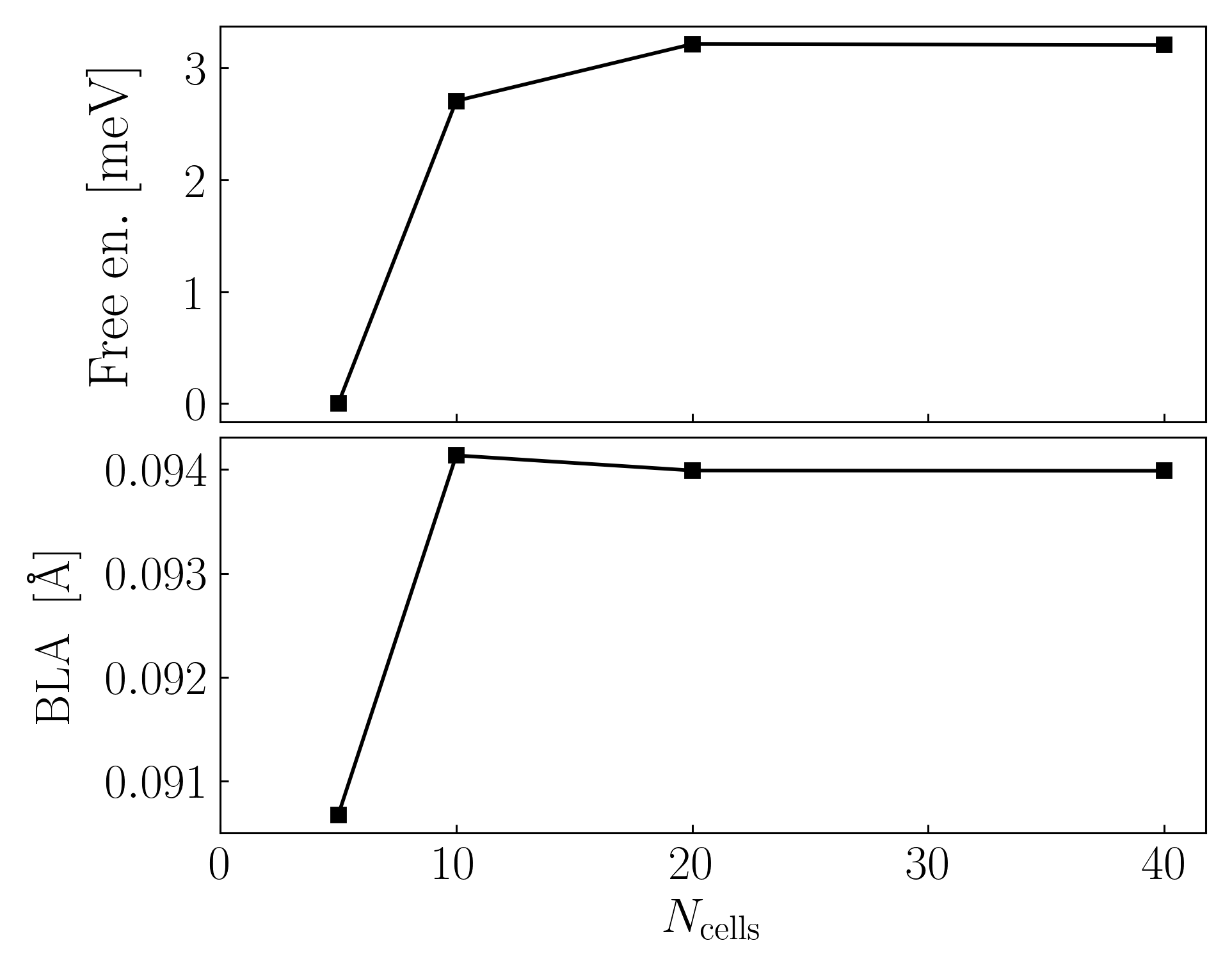} 
    \caption{For the exemplary case with $\Delta=0$ and at $T=0~\mathrm{K}$, we display the convergence of the minimized free energy and of the BLA obtained from the minimized density distribution, with respect to the number $N_\mathrm{cells}$ of elementary diatomic cells in the supercells.}
	\label{fig:SSCHA_convergence}
    \end{minipage}
\end{figure}

All the SSCHA minimizations were performed using $N_\mathrm{conf}=4000$ supercell configurations. In Figure \ref{fig:SSCHA_convergence}, for a representative case with $\Delta=0$ and $T=0~\mathrm{K}$, we show the convergence of the optimal free energy and of the optimal BLA obtained from the minimized density distributions, with respect to the number $N_\mathrm{cells}$ of elementary diatomic cells in the $N_\mathrm{conf}$ supercells. For consistency with respect to the calculations performed to tune the parameters of the model, where we used $N_\mathrm{cells}=1$ and $N_k=320$, here we used $N_k=320/N_\mathrm{cells}$ to obtain the results displayed in the Figure. We then chose $N_\mathrm{cells}=40$, and hence $N_k=8$, for all the SSCHA minimizations presented in the main text. 

\section{Details on the calculation of the piezoelectric coefficients} \label{app_piezo}
\begin{figure}[!htb]
    \begin{minipage}[c]{1.0\linewidth}
	\centering
        \includegraphics[width=1.0\textwidth]{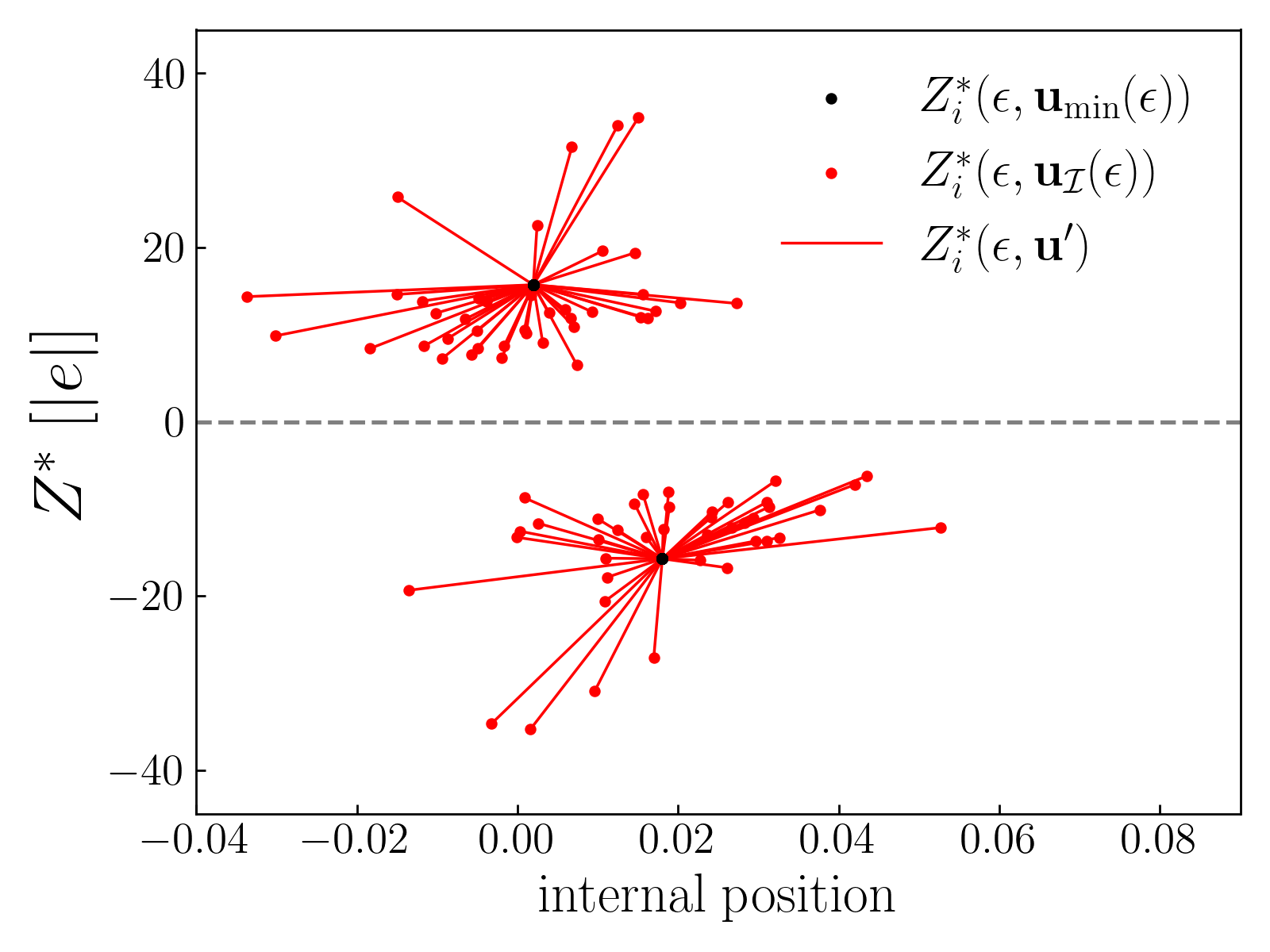} 
        \caption{For one of the $N_\mathrm{conf}$ configurations of a generic $\Delta\neq0$, the red lines show the behaviour of the Born effective charges of the $N_\mathrm{at}=80$ atoms in the supercell, computed along the path $\boldsymbol{u}'=\lambda\boldsymbol{u}_\mathcal{I}(\epsilon) + (1-\lambda)\boldsymbol{u}_\mathrm{min}(\epsilon)$, with $\lambda\in[0,1]$. In particular, each line illustrates the values of $Z_i^*(\epsilon,\boldsymbol{u}^\prime)$ of atom $i$, for $i=1,\dots,N_\mathrm{at}$, and is displayed against the values of the fractional coordinates ${u}_i^\prime$. The black dots are the two values of the effective charges computed on the supercell with atoms in positions $\boldsymbol{u}_\mathrm{min}(\epsilon)$. Since this supercell is made by the repetition of $N_\mathrm{cells}$ unit cells with atoms in positions $\boldsymbol{u}_\mathrm{min}^\mathrm{u.c.}(\epsilon)$, we observe just two different values, one for all the atoms with onsite energy difference $+\Delta$, and the other for all the atoms with $-\Delta$. The red dots are the values of the effective charges computed on the supercell with atoms in positions $\boldsymbol{u}_\mathcal{I}(\epsilon)$. }
	\label{fig:Zeff_vs_u}
    \end{minipage}
\end{figure}

\begin{figure}[!htb]
    \begin{minipage}[c]{1.0\linewidth}
	\centering
        \includegraphics[width=1.0\textwidth]{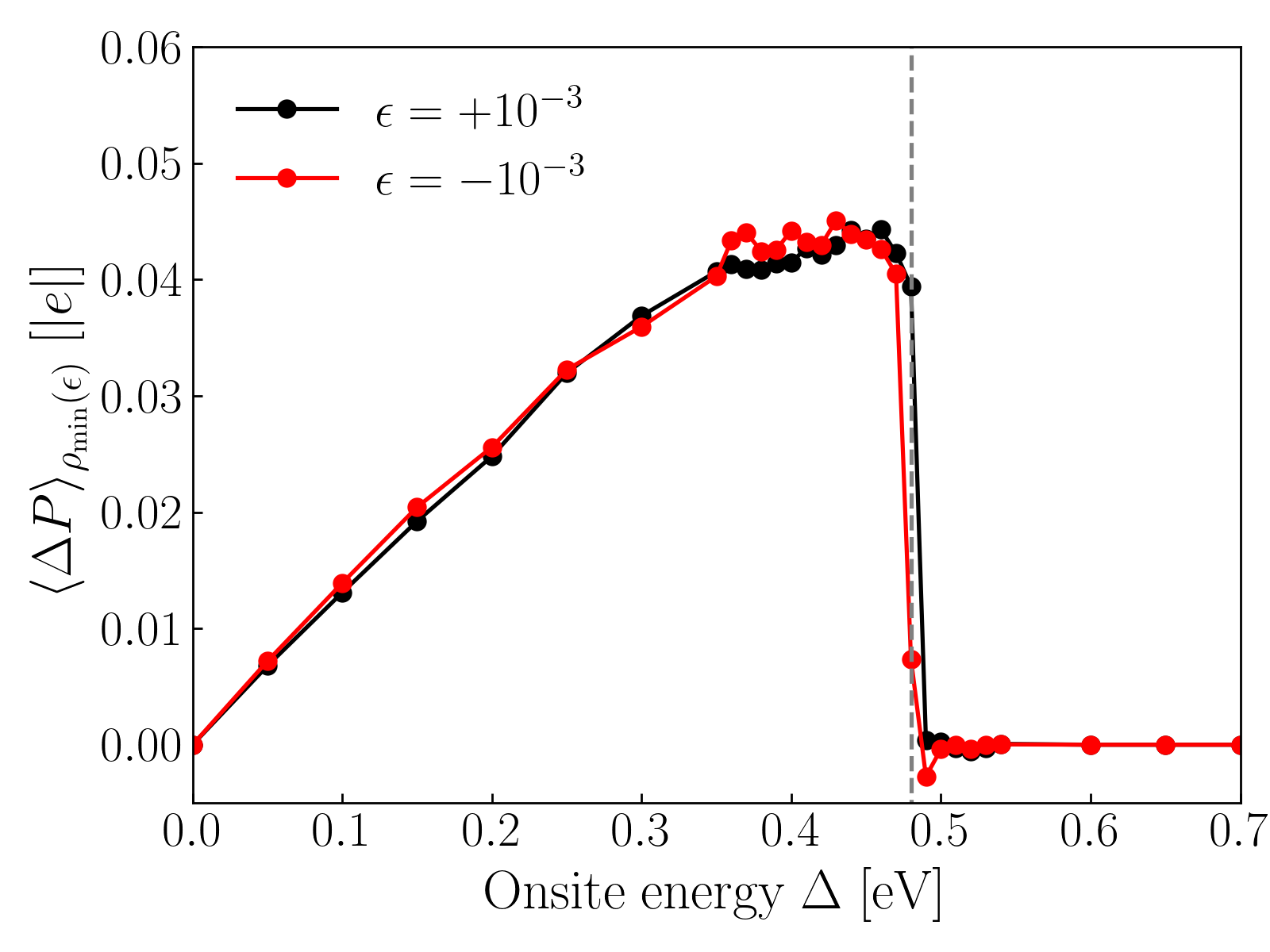} 
        \caption{For the two values of finite strain $\epsilon=\pm10^{-3}$ applied to the chain, we display the values of $\langle \Delta P \rangle_{\rho_\mathrm{min}(\epsilon)}$ obtained from Equation (\ref{eq:DeltaP_avg}) for values of the onsite energy difference $\Delta$.}
	\label{fig:DeltaP_vs_D}
    \end{minipage}
\end{figure}

In the absence of fluctuations, we considered a single diatomic unit cell ($N_\mathrm{cells}=1$) in periodic boundary conditions with $N_k=320$ and we computed the piezoelectric coefficient $c_\mathrm{piezo}$ as finite differences of the dipole moment per unit cell length $P$ with respect to a strain $\epsilon$, namely
\begin{equation}
    \label{eq:c_piezo_RM}
    c_\mathrm{piezo} = \frac{P(+\epsilon,\boldsymbol{u}(+\epsilon)) - P(-\epsilon,\boldsymbol{u}(-\epsilon))}{2 \epsilon},
\end{equation}
where we obtained $P$ with the Berry phase approach\cite{berry1984quantal,resta1994macroscopic} on strained diatomic unit cells with $\epsilon=\pm10^{-3}$. The fractional coordinates $\boldsymbol{u}(\pm\epsilon)=(u_1(\pm\epsilon),u_2(\pm\epsilon))$ used for the calculations were relaxed to minimize the total energy per unit cell of the strained chain, following the procedure described in Ref~\cite{villani2024giant}. 
To compute the values of $c_\mathrm{piezo}(\boldsymbol{R}_\mathrm{min})$, we proceeded similarly as for the calculation of $Z^*(\boldsymbol{R}_\mathrm{min})$, and considered a single diatomic unit cell ($N_\mathrm{cells}=1$) with the atoms $1$ and $2$ in positions $\langle r_1 \rangle_{\rho_\mathrm{min}(\pm\epsilon)}$ and $\langle r_2 \rangle_{\rho_\mathrm{min}(\pm\epsilon)}$, respectively. Then, as in Equation (\ref{eq:c_piezo_RM}), we computed finite differences of polarization  
\begin{align}
    c_\mathrm{piezo}(\boldsymbol{R}_\mathrm{min}) = \frac{ P(+\epsilon,\boldsymbol{u}_\mathrm{min}^\mathrm{u.c.}(+\epsilon)) - P(-\epsilon,\boldsymbol{u}^\mathrm{u.c.}_\mathrm{min}(-\epsilon))}{2 \epsilon}
\end{align}
where $\boldsymbol{u}_\mathrm{min}^\mathrm{u.c}(\pm\epsilon) = ( \langle r_1 \rangle_{\rho_\mathrm{min}(\pm\epsilon)}/a(\pm\epsilon), \langle r_2 \rangle_{\rho_\mathrm{min}(\pm\epsilon)}/a(\pm\epsilon) )$ indicates the fractional coordinates of the atoms in the unit cell, 
$P$ is computed with the Berry phase approach on the unit cells, and we used $\epsilon=\pm10^{-3}$. 
We defined the quantum-anharmonic corrected piezoelectric coefficient $\langle c_\mathrm{piezo} \rangle_{\rho_\mathrm{min}}$ of the elementary diatomic cell as
\begin{align}
    \label{eq:def_cpiezo_avg}
    \langle c_\mathrm{piezo} \rangle_{\rho_\mathrm{min}} = \frac{\langle P \rangle_{\rho_\mathrm{min}(+\epsilon)} - \langle P \rangle_{\rho_\mathrm{min}(-\epsilon)}}{2 \epsilon},
\end{align}
where we defined the unit cell's quantum-anharmonic-corrected dipole moment per cell length as 
\begin{equation}
    \label{eq:P_avg}
    \langle P \rangle_{\rho_\mathrm{min}(\epsilon)} \simeq \frac{1}{N_\mathrm{cells}N_\mathrm{conf}}  \sum_{\mathcal{I}=1}^{N_\mathrm{conf}} P (\epsilon,\boldsymbol{u}_\mathcal{I}(\epsilon)) 
\end{equation}
where $\boldsymbol{u}_\mathcal{I}(\epsilon) = \boldsymbol{R}_\mathcal{I}(\epsilon)/a(\epsilon) =(u_{\mathcal{I},1}(\epsilon),\dots,u_{\mathcal{I},N_\mathrm{at}}(\epsilon))$ is the collection of the atomic fractional coordinates of all the $N_\mathrm{at}$ atoms in the supercell of the $\mathcal{I}$-th configuration and $\rho_\mathrm{min}(\epsilon,\boldsymbol{u}_\mathcal{I}(\epsilon))$ is the probability associated to $\boldsymbol{u}_\mathcal{I}(\epsilon)$. 
We highlight that, in above Equation (\ref{eq:P_avg}), $P$ is the dipole moment of the supercell divided by the length of an elementary cell, so it requires calculating the polarization of each supercell. To avoid doing so, we exploit the following identity:
\begin{equation}
    P (\epsilon,\boldsymbol{u}_\mathcal{I}(\epsilon)) = \Delta P_\mathcal{I}(\epsilon) + P(\epsilon,\boldsymbol{u}_\mathrm{min}(\epsilon)), 
\end{equation}
where we defined
\begin{equation}
    \Delta P_\mathcal{I}(\epsilon) = P (\epsilon,\boldsymbol{u}_\mathcal{I}(\epsilon)) - P(\epsilon,\boldsymbol{u}_\mathrm{min}(\epsilon)), 
\end{equation}
and $\boldsymbol{u}_\mathrm{min}(\epsilon)=\boldsymbol{R}_\mathrm{min}/a(\epsilon)$ is the vector with the fractional coordinates of the $N_\mathrm{at}$ atoms in the optimal configuration obtained from the SSCHA minimizations. Exploiting the fact that $P(\epsilon,\boldsymbol{u}_\mathrm{min}(\epsilon))$ is independent on $\mathcal{I}$, we rewrite Equation (\ref{eq:P_avg}) as follows
\begin{align}
    \langle P \rangle_{\rho_\mathrm{min}(\epsilon)} &= \langle \Delta P \rangle_{\rho_\mathrm{min}(\epsilon)} + \frac{P(\epsilon,\boldsymbol{u}_\mathrm{min}(\epsilon))}{N_\mathrm{cells}} 
\end{align}
where
\begin{equation}
    \label{eq:DeltaP_avg}
    \langle \Delta P \rangle_{\rho_\mathrm{min}(\epsilon)} \simeq \frac{1}{N_\mathrm{cells}N_\mathrm{conf}} \sum_{\mathcal{I}=1}^{N_\mathrm{conf}} \Delta P _\mathcal{I}(\epsilon) .
\end{equation}
This allows us to rewrite the quantum-anharmonic-corrected piezoelectric coefficient $\langle c_\mathrm{piezo} \rangle_{\rho_\mathrm{min}}$ defined in Equation (\ref{eq:def_cpiezo_avg}) as
\begin{align}
    \langle c_\mathrm{piezo} \rangle_{\rho_\mathrm{min}} 
    &=
    \frac{\langle \Delta P \rangle_{\rho_\mathrm{min}(+\epsilon)} - \langle \Delta P \rangle_{\rho_\mathrm{min}(-\epsilon)}}{2\epsilon} + \nonumber \\ &+ \frac{1}{N_\mathrm{cells}}\frac{P(+\epsilon,\boldsymbol{u}_\mathrm{min}(+\epsilon)) - P(-\epsilon,\boldsymbol{u}_\mathrm{min}(-\epsilon))}{2\epsilon}.
\end{align}
Because $P(\epsilon,\boldsymbol{u}_\mathrm{min}(\epsilon))$ is defined over a supercell made by $N_\mathrm{cells}$ repetitions of the unit cells with atoms in positions $\boldsymbol{u}_\mathrm{min}^\mathrm{u.c.}(\epsilon)$, it holds $P(\epsilon,\boldsymbol{u}_\mathrm{min}(\epsilon)) / N_\mathrm{cells} = P(\epsilon,\boldsymbol{u}_\mathrm{min}^\mathrm{u.c.}(\epsilon))$, allowing us to write:  
\begin{equation}
    \label{eq:c_piezo_avg}
    \langle c_\mathrm{piezo} \rangle_{\rho_\mathrm{min}}  = \frac{\langle \Delta P \rangle_{\rho_\mathrm{min}(+\epsilon)} - \langle \Delta P \rangle_{\rho_\mathrm{min}(-\epsilon)}}{2\epsilon} + c_\mathrm{piezo}(\boldsymbol{R}_\mathrm{min}) ,
\end{equation}
where we isolated the contribution $c_\mathrm{piezo}(\boldsymbol{R}_\mathrm{min})$ from the contribution due to the average over quantum fluctuations. To obtain this latter term, we computed the values of $\Delta P_\mathcal{I}(\epsilon)$ entering in Equation (\ref{eq:DeltaP_avg}) as follows
\begin{equation}
    \label{eq:DeltaP_I_integral}
    \Delta P_\mathcal{I}(\epsilon) = \int_{\boldsymbol{u}_{\mathrm{min}}(\epsilon)}^{\boldsymbol{u}_{\mathcal{I}}(\epsilon)} \boldsymbol{Z}^*(\epsilon,\boldsymbol{u}') \cdot \dd \boldsymbol{u}',
\end{equation}
where $\boldsymbol{Z}^*(\epsilon,\boldsymbol{u}')=(Z_1^*(\epsilon,\boldsymbol{u}'),\dots,Z_{N_\mathrm{at}}^*(\epsilon,\boldsymbol{u}'))$ is a vector with the Born effective charges of the $N_\mathrm{at}$ atoms in the supercell, computed using linear response theory on a chain with an applied finite strain $\epsilon$ and with atoms in positions $\boldsymbol{u}'$. The integral can be taken along any path from the vector $\boldsymbol{u}_{\mathrm{min}}(\epsilon)$ to the vector $\boldsymbol{u}_{\mathcal{I}}(\epsilon)$. As shown in Figure \ref{fig:Zeff_vs_u} for an exemplary configuration, the behaviour of $\boldsymbol{Z}^*(\epsilon,\boldsymbol{u}')$ is approximately linear with respect to $\boldsymbol{u}'$ along the path $\boldsymbol{u}'(\lambda)=\lambda\boldsymbol{u}_\mathcal{I}(\epsilon) + (1-\lambda)\boldsymbol{u}_\mathrm{min}(\epsilon)$, with $\lambda\in[0,1]$. This allows us to compute the integral as follows:
\begin{align}
    \label{eq:Zeff_integral}
    \int_{0}^{1} & \boldsymbol{Z}^*(\epsilon,\boldsymbol{u}'(\lambda)) \cdot \left[ \boldsymbol{u}_{\mathcal{I}}(\epsilon) - \boldsymbol{u}_\mathrm{min}(\epsilon) \right] \dd \lambda \simeq \nonumber \\ & \simeq \frac{1}{2}\left[ \boldsymbol{Z}^*(\epsilon,\boldsymbol{u}_{\mathcal{I}}(\epsilon)) +  \boldsymbol{Z}^*(\epsilon,\boldsymbol{u}_\mathrm{min}(\epsilon)) \right] \cdot \left[ \boldsymbol{u}_{\mathcal{I}}(\epsilon) - \boldsymbol{u}_\mathrm{min}(\epsilon) \right].
\end{align}
As shown in Figure \ref{fig:DeltaP_vs_D} and consistently with the results shown in Figure \ref{fig:piezo_comparison_QAE}, the contribution of $\langle \Delta P \rangle_{\rho_\mathrm{min}(\epsilon)}$ to the quantum-anharmonic-corrected piezoelectric coefficient is negligible. 
This result contains further insight on the effects of quantum anharmonicity on the polar response. Indeed, thanks to Equation (\ref{eq:Zeff_integral}), we express $\langle \Delta P \rangle_{\rho_\mathrm{min}(\epsilon)}$ in terms of the covariance between atoms' effective charges and their internal positions expressed in fractional coordinates. Plugging Equations (\ref{eq:DeltaP_I_integral}) and (\ref{eq:Zeff_integral}) into Equation (\ref{eq:DeltaP_avg}), we have 
\begin{equation}
    \langle \Delta P \rangle_{\rho_\mathrm{min}(\epsilon)} = \frac{1}{2} \left[ \langle \boldsymbol{Z}^*\cdot \boldsymbol{u} \rangle_{\rho_\mathrm{min}(\epsilon)} - \langle \boldsymbol{Z}^*\rangle_{\rho_\mathrm{min}(\epsilon)} \cdot \langle \boldsymbol{u} \rangle_{\rho_\mathrm{min}(\epsilon)} \right],
\end{equation}
where we defined
\begin{equation}
    \label{eq:def_Zeff_u_QAE}
    \langle \boldsymbol{Z}^*\cdot \boldsymbol{u} \rangle_{\rho_\mathrm{min}(\epsilon)} \simeq \frac{1}{N_\mathrm{cells}N_\mathrm{conf}}\sum_{\mathcal{I}=1}^{N_\mathrm{conf}} \boldsymbol{Z}^*(\epsilon,\boldsymbol{u}_\mathcal{I}(\epsilon))\cdot\boldsymbol{u}_\mathcal{I}(\epsilon),
\end{equation}
and we indicated with $\langle \boldsymbol{Z}^* \rangle_{\rho_\mathrm{min}(\epsilon)}=(\langle Z^*_1 \rangle_{\rho_\mathrm{min}(\epsilon)}, \langle Z^*_{2} \rangle_{\rho_\mathrm{min}(\epsilon)})$ a vector with the averaged values of the effective charges of the $2$ atoms in the diatomic unit cell, computed as
\begin{equation}
    \langle Z^*_i \rangle_{\rho_\mathrm{min}(\epsilon)} \simeq \frac{1}{N_\mathrm{cells}N_\mathrm{conf}}\sum_{n=1}^{N_\mathrm{cells}} \sum_{\mathcal{I}=1}^{N_\mathrm{conf}} {Z}^*_{2n-i}(\epsilon,\boldsymbol{u}_\mathcal{I}(\epsilon)),
\end{equation}
and it holds $\langle \boldsymbol{u} \rangle_{\rho_\mathrm{min}(\epsilon)} = \boldsymbol{u}_\mathrm{min}^\mathrm{u.c.}(\epsilon)$. Since $\langle \Delta P \rangle_{\rho_\mathrm{min}} \simeq 0$, it holds 
\begin{equation}
    \langle \boldsymbol{Z}^*\cdot \boldsymbol{u} \rangle_{\rho_\mathrm{min}} \simeq \langle \boldsymbol{Z}^*\rangle_{\rho_\mathrm{min} } \cdot \langle \boldsymbol{u} \rangle_{\rho_\mathrm{min}},
\end{equation}
which in the context of linear-response theory within the SSCHA framework\cite{monacelli2021time,siciliano2023wigner} implies negligible two-phonons effects. We also used this relation to compute the internal-relaxation contribution $\langle c_\mathrm{piezo}^\mathrm{i.r.} \rangle_{\rho_\mathrm{min}}$ averaged over the ionic fluctuations as 
\begin{align}
    \langle c_\mathrm{piezo}^\mathrm{i.r.} \rangle_{\rho_\mathrm{min}}
    &= \langle \sum_i Z^*_i \frac{\partial u_i}{\partial \epsilon} \rangle_{\rho_\mathrm{min}} \\
    \label{eq:c_piezo_ir_avg}
    & \simeq \sum_i \langle Z^*_i \rangle_{\rho_\mathrm{min}} \langle \frac{\partial u_i }{\partial \epsilon} \rangle_{\rho_\mathrm{min}} \\
    & \simeq \sum_i \langle Z^*_i \rangle_{\rho_\mathrm{min}} \frac{\langle \boldsymbol{u} \rangle_{\rho_\mathrm{min}(+\epsilon)} - \langle \boldsymbol{u}\rangle_{\rho_\mathrm{min}(-\epsilon)}}{2\epsilon}.
\end{align}
All the values of quantum-anharmonic-corrected averages $\langle \bullet \rangle$ of the quantities defined in this Appendix were computed with $N_\mathrm{conf}=504$, both here and in the main text. 

\end{document}